\documentclass[fleqn,usenatbib]{mnras}

\usepackage{newtxtext,newtxmath}
\usepackage[T1]{fontenc}
\usepackage{ae,aecompl}
\usepackage{enumerate}
\usepackage[switch, modulo]{lineno}
\usepackage[toc,page]{appendix}
\usepackage{hyperref}
\usepackage{multirow}

\usepackage{xcolor}

\usepackage{graphicx}	
\usepackage{bm}

\usepackage{graphicx}
\usepackage{bm}

\usepackage{lineno}
\usepackage{subcaption}

\def\calMa{{V_{4}}}

\def\ba{\begin{array}}
\def\ea{\end{array}} 
\def\bea{\begin{eqnarray}}
\def\eea{\end{eqnarray}}
\def\beq{\begin{equation}}
\def\eeq{\end{equation}}
\def\ben{\begin{enumerate}}
\def\een{\end{enumerate}}

\def\brr{\begin{array}}
\def\err{\end{array}}

\def\rt{\tilde{r}}

\def\calMa{{V_{4}}}

\title[Cold Collapse and Bounce of an FLRW Cloud]
{Cold Collapse and Bounce of an FLRW Cloud}

\author[S. Pradhan, M. Gabler \& E. Gazta\~naga]{
Swaraj Pradhan$^{1,2}$\thanks{E-mail: \href{000jaraws@gmail.com }{000jaraws@gmail.com }}, Michael Gabler$^3$ \& Enrique Gazta\~naga$^{1,4,5}$ 
\\
$^1$ Institute of Cosmology and Gravitation, University of Portsmouth, Portsmouth, PO1 3FX, UK\\
$^2$ Department of Physical Sciences, Indian Institute of Science Education and Research Kolkata, Mohanpur 741246, India \\
$^3$Departamento de Astronomía y Astrofísica, Universitat de València, E-46100 Burjassot (València), Spain\\
$^4$ Institute of Space Sciences (ICE, CSIC), 08193 Barcelona, Spain \\
$^5$ Institut d\'~Estudis Espacials de Catalunya (IEEC), 08034 Barcelona, Spain \\
}

\begin{document}
\label{firstpage}
\pagerange{\pageref{firstpage}--\pageref{lastpage}}
\maketitle

\begin{abstract}
We study the collapse of spherical cold clouds beyond black hole formation to investigate the possibility of a bounce in the in-falling matter when a critical density or pressure is reached. As a first step, we analyse the pressureless collapse in general relativity (GR), where an analytic solution exists, and demonstrate that an equivalent Newtonian solution can be derived. Such equivalence also holds for spherically symmetric perfect fluids with uniform density and non-vanishing pressure.
We numerically investigate the Newtonian collapse of such clouds with masses of $5$, $20$, and $1000$ M$_\odot$ obeying a polytropic equation of state (EoS). By choosing EoS parameters inspired by typical neutron star conditions, we observe bounces at and above nuclear saturation density. Assuming approximate uniformity, we explore the equivalent GR behaviour of the matter during the bounce. 
Our findings are as follows:
(i) A GR bounce occurs around the ground state of the matter, characterized by $P = -\rho$.  
(ii) The GR solution differs significantly from the Newtonian result due to the presence of curvature ($k \neq 0$).  
(iii) Both the curvature and the ground state are crucial factors in allowing a GR bounce to occur.

\end{abstract}


\maketitle

\section{Introduction}
\label{S:1}
In 1970, Hawking and Penrose proved that singularities cannot be avoided in general relativity (GR) \citep{Hawking_Penrose_Singularity}. Due to the success of these singularity theorems, the solutions and ideas of non-singular or bouncing universes have been overlooked in the last 30 yr of the 20th century \citep{Brandenberger_bouncing}. A bounce of collapsing matter at supranuclear densities naturally avoids the formation of a singularity. To allow for such a bounce in this class of solutions, Einstein's gravity would have to violate the strong energy conditions (SEC). For a perfect fluid with energy density $\rho$ and pressure $P$, the SEC corresponds to $\rho+3P \ge 0$. This SEC violation only started to be accepted in the arising models of inflation \citep{Burd:1988ss_SEC, Barrow_SEC}. The discovery of positive acceleration of the Universe \citep{Riess_1998, Schmidt_1998, Perlmutter_1999, Jimenez_2002,2003ApJ...597L..89F, Eisenstein_2005, Betoule_2012} proved the existence of SEC violation, and the models of bouncing cosmologies received new interest \citep{Brandenberger_bouncing}. These models aim to solve the problems encountered by the standard model of cosmology, among which are: (i) the singularity problem: the existence of a singularity means that the space-time is geodesically past incomplete; (ii) the flatness problem: the critical value of matter\textendash energy density of the Universe needs to be extremely fine-tuned, any small deviation from this value will lead to a non-flat universe; and (iii) the horizon problem: we see a uniform cosmic microwave background (CMB) sky where superhorizon scales appear (these are scales larger than the Hubble horizon at the time when the CMB photons were emmitted). How is this possible if the structures were not in casual contact in the past? Alternatively, new theories of modified gravity and the concept of quantum gravity \citep{Rovell_QG, Olmedo_QG} were developed. However, the latter is incomplete and not readily testable \citep[see][for a review]{bounce_rev1}.

One realization of the bouncing universe is the {\it Black Hole (BH) Universe} or BHU model presented in \cite{Gaztanaga:bhu1,Gaztanaga:bhu2}. In this model, some of the problems faced by the standard model are avoided without the need to introduce new theories of modified gravity or Dark Energy (DE). In the BHU model, our observable Universe is in the interior of its corresponding BH event horizon, and the $\Lambda$-term in the standard cosmology arises as a boundary to the evolution of the system inside the BH event horizon \citep{Gaztanaga:2021lamb1, Gaztanaga:2022lamb2}. The initial condition of such a universe could be a tenuous, cold, but large Friedmann\textendash Lemaitre\textendash Robertson\textendash Walker (FLRW) cloud containing the total mass of our observable Universe. This cloud collapses under its own gravity into a BH. For such a large mass, the density at the time of the formation of the BH is very small, and the pressureless collapse inside the BH event horizon proceeds as if there was no BH formed. Note that we use the expression {\it cold} to describe a fluid at zero temperature, resulting in zero thermal pressure. We use both expressions equally. Therefore, if not stated otherwise, {\it pressureless} refers to zero thermal pressure. 

The gravitational collapse occurs homologously on a shorter time scale 
than thermal interaction of particles could create any significant  thermal pressure.
Thus, it is reasonable to assume that the fluid of the cloud stays cold during the entire collapse process. This means that if nothing else occurs, the density of the cloud continues to increase during the collapse until a singularity is formed.

However, there are effects that can slow down, if not even bring the cold collapse to a halt. Whenever one of the fermionic constituents of the cloud reaches its quantum ground state, the Pauli exclusion principle creates a degeneracy pressure independent of temperature. Assuming a cloud of neutral hydrogen, first the electrons would become degenerate and create a pressure that could stabilize configurations up to the Chandrasekhar mass limit ($\sim1.4$ M$_\odot$). Typical maximum densities allowed in this way are lower than $\rho\lesssim10^7 $g cm$^{-3}$. At higher densities, reached for larger initial masses, the electrons recombine with protons, and the collapse continues. Similarly, at nuclear saturation densities, the collapse could be stopped due to the quantum pressure of the ground state of nuclear matter. The maximum mass allowed for this to happen sensitively depends on the equation of state (EoS) of nuclear matter \citep{Hebeler_2013, Ferreira2021}, but it should be between $2$ and $3$ M$_\odot$.

Similar scenarios are actually observed in nature at the end of the life of massive stars. The fate of these objects depends on the final mass of their collapsing cores. Stars can lose a significant amount of their initial mass during their different burning stages through stellar winds or at the end of their lifetime by ejecting the outer layers during the collapse of the central cores. For stars with zero-age main-sequence masses of up to about $\sim8$ M$_\odot$, the final cores reach at most the Chandrasekhar mass ($\sim1.4$ M$_\odot$). Therefore, these stars end their lives as White Dwarfs. The cores of more massive stars ($\sim8$ M$_\odot\lesssim M \lesssim40$ M$_\odot$) eventually become unstable and collapse once the Chandrasekhar mass is reached. This collapse is stopped once the collapsing core of the progenitor of $\sim1.4$ M$_\odot$ reaches nuclear saturation density, and matter cannot be compressed further. The still-infalling outer layers bounce back from the proto-neutron star surface and create an outward-moving shock wave. However, this shock wave stalls because of energy losses by neutrinos and dissociation of the infalling nuclei. The details of how this shock revives and the bounce finally leads to an exploding supernova are a fascinating topic of current astrophysical research \citep[see][for reviews]{bethe1990,burrows1995,janka2012,foglizzo2017,janka2017,limongi2017}. What is of interest to the current work is the existence of the bounce of the infalling matter. Note that the densities in neutron stars and in the nucleus of an atom are comparable, despite the former having a mass \( 10^{57} \) times larger than the latter. Having this in mind, the step from stellar size objects we study here with tens to thousand solar masses to the mass of the Universe ($\sim10^{22}$ M$_\odot$; \citealt{gaztanaga_mou}) appears less exaggerated.

However, we have to note that there are essential differences between the cold collapse we are interested in and a supernova explosion. Supernova are `hot' events. This not only means that there are photons present that mediate thermal pressure but also that there are plenty of nuclear reactions occurring simultaneously. Depending on the conditions of temperature, pressure, and density, different reactions create and destroy nuclei, releasing copious amounts of neutrinos that may interact with matter in different locations. The outward-moving shock increases entropy and, hence, the temperature of the infalling matter, which leads to the dissociation of the nuclei into nucleons. The composition of matter depends sensitively on all the conditions and on the dynamics of the entire process. Fortunately, most of these complications disappear once we make the reasonable assumption of a cold collapse of neutral hydrogen. 

The scenarios discussed above describe what happens for density distributions whose total mass is sufficiently low, and the bounce occurs before the respective distribution is confined within its gravitational radius. The larger the total mass, the lower the density at which a BH forms, and consequently, for more massive density distributions, (supra)nuclear densities only appear inside the corresponding horizon. This brings us back
to our BHU model, where we investigate the collapse of a very massive cloud, for which the cold collapse occurs within the black hole itself. Given the enormous mass, one might assume that the collapse would not stop at nuclear saturation density, potentially leading to even higher densities. 

In the BHU model (\citealt{gaztanaga_mou}), the working hypothesis is that some kind of quantum degeneracy pressure or a more energetic ground state of matter could be reached at some supranuclear density. For masses as large as our Universe, the threshold density could be much larger than the nuclear saturation density. In analogy to a supernova, the matter then may not be compressed beyond this threshold, the collapse halts abruptly, and the infalling matter bounces back. Consequently, the propagation of matter changes from collapse to expansion, potentially providing an alternative explanation for cosmic inflation and the related dark energy. A mathematical singularity, which is inevitable in non-bouncing GR models, can be avoided with such a bounce. After the big bang, the superhorizon perturbations from the collapsing phase re-enter the horizon as the expansion proceeds and seed structure formation (\citealt{Gaztanaga:bhu}), solving the horizon problem. As we see in the later sections, the curvature value in the GR framework ($k$) goes to a finite value during the bounce, but the space\textendash time flattens out as the expansion proceeds and density reduces. Therefore, the flatness problem is solved in a manner similar to that of inflation.

As a first step towards a model for our Universe, we first need to understand the mechanism of the cold collapse and bouncing. We begin our study by considering masses in the range of $5-1000$ M$_\odot$, where numerical results of the formation of BH are available. Traditionally, simulations done in GR usually (i) stop when a trapped surface or an apparent horizon forms \citep[e.g.][]{Escriva:spribhos}, (ii) `excise' the central part of the evolution inside the Apparent Horizon \citep{Seidel_excision, Marsa_Choptuik_excision, Alcubierre_2001, Duez2004_excision, Montero_2012}, or (iii) replace the BH interior with analytic solutions, for example, `punctures' \citep{Campanelli_2006_puncture, Baker_2006_puncture, Brugmann_puncture}. What happens inside the black hole is generally not of interest, since no information can come out of the black hole and does not influence the exterior dynamics. By ignoring the interior evolution, the singularity arising from the BH formation is effectively excluded, and the exterior evolution beyond the BH formation can be followed. 

In contrast, here we are interested in what happens when the matter continues to collapse inside the BH. 
\cite{Gaztanaga:bhu1} showed how a spherically symmetric FLRW cloud collapses to form a BHU (called the Black Hole Universe). The collapse continues inside the BH. This is the starting point of our study, and we investigate the dynamics of the GR solution for the collapse of such a uniform, pressure-less spherical mass distribution with total mass $M$. For this configuration, the relativistic solution is mathematically equivalent to the corresponding Newtonian solution for a system with net zero total energy. We further assert that this kind of mathematical equivalence exists also for cases with non-zero pressure. To investigate the collapses and bounces of uniform spherical clouds of different masses quantitatively, we simulate their evolution numerically in
Section \ref{sec:simus}. For simplicity, we assume a polytropic EoS. In particular, we investigate the initial conditions and EoS parameters required for the bounce and the properties of the latter. We conclude in Section \ref{sec:conclusions}.

\section{Collapse of a uniform spherical mass distribution}
\label{S:2}
To find the analytical equations that describe the dynamics of the collapse of a uniform spherical mass distribution with total mass $M$, we closely follow the approach presented in \cite{gaztanaga_mou}.
The spherically symmetric metric $\boldsymbol{g}_{\mu\nu}$ in polar coordinates with a proper time $\text{d}\boldsymbol{x}^\mu =(\rm d\tau,\rm d\chi,\rm d\theta,\rm d\phi)$ can be expressed by the line element \citep{Tolman1934,Oppenheimer-Snyder,Misner-Sharp}:
\beq
 \text{d}s^2 = \boldsymbol{g}_{\mu\nu} \text{d}\boldsymbol{x}^\mu \text{d}\boldsymbol{x}^\nu =
- \text{d}\tau^2 + e^{\lambda(\tau,\chi)} \text{d}\chi^2 + r^2(\tau,\chi) \text{d}\Omega^2 .
\label{eq:SphericalMetric2}
\eeq
Here, we use units of speed of light $c=1$, and the time and radial coordinates $(\tau,\chi)$ are comoving with the matter (we use $t$ to refer to rest frame or Newtonian coordinates). The corresponding metric is sometimes reffered to as the Lemaitre\textendash Tolman \citep{Lemaitre,Tolman1934} or the Lemaitre\textendash Tolman\textendash Bondi metric. 
This metric is localized around a central reference point in space, which we have set as the origin ($\vec{r}=0$) for simplicity.

For an observer moving with a perfect fluid, the energy\textendash momentum tensor is diagonal:
$\boldsymbol{T}_\mu^{~\nu} = diag[-\rho,P,P,P]$,
where $\rho=\rho(\tau,\chi)$ is the relativistic energy density and $P=P(\tau,\chi)$ is the pressure.
The off-diagonal terms of the field equation vanish, for example, $8\pi G \boldsymbol{T}^{~1}_0=\boldsymbol{G}^{~1}_0=0$. These equations can be reformulated into:
\beq
\left(\upartial_\tau \lambda\right) (\upartial_\chi r) = 2 \upartial_\tau (\upartial_\chi r)\,.
\eeq 
This equation can be solved with the ansatz: $e^\lambda = C (\upartial_\chi r)^2 $, where
$\upartial_\tau C=0$. The particular case $C=1$ corresponds to a flat geometry:
\beq
\text{d}s^2 = - \text{d}\tau^2 + 
\left(\upartial_\chi r\right)^2 \text{d}\chi^2 + r^2 \text{d}\Omega^2\,.
\label{eq:LBT}
\eeq
Consequently, there is only one function required to be solved: $r=r(\tau,\chi)$ and $r$ corresponds to the radial proper distance. For a perfect fluid in the comoving frame:
\beq
\boldsymbol{T}_\mu^{~\nu} = diag[-\rho,P,P,P] = \left( \begin{array}{cccc} 
-\rho & 0 & 0 & 0 \\
0 &  P & 0 & 0 \\
0 & 0 &  P & 0 \\
0 & 0 & 0 &  P \\\end{array} \right)\,.
\label{eq:Tdiag}
\eeq
The field equations for $r$  are then:
\bea
H^2 &\equiv&  r_H^{-2} \equiv \left( \frac{\dot r}{r} \right)^2 = \frac{2 G M(\tau,\chi)}{r^3} \label{eq:H2} ,
\\ M(\tau,\chi)  &\equiv& \int_0^\chi \rho \,   4\pi r^2 \left(\upartial_\chi r\right)  \text{d}\chi\,,
\label{eq:Mass}
\eea
where $M$ is the mass inside $\chi$ at time $\tau$ and the dot here is $\dot{r} \equiv \upartial_\tau r$. See Appendix \ref{app:mass} for a discussion on the Newtonian and relativistic masses.
The remaining field equations correspond to energy conservation, which we discuss later in equation\,\eqref{eq:parta0}.
When $\rho=\rho(\tau)$ is uniform, we have $M={4\pi} r^3 \rho / 3$ 
and:
\beq
r=a(\tau) \chi ,
\label{eq:r=achi}
\eeq
where $a(\tau)$ denotes the scale factor of the collapsing cloud. Thus, the metric in equation\,\eqref{eq:LBT} reproduces  the flat (global) FLRW metric: 
\beq
\text{d}s^2=-\text{d}\tau^2 + a^2 \left(\text{d}\chi^2 + \chi^2 \text{d}\Omega^2\right)\,.
\label{eq:FLRW}
\eeq
Equation\,\eqref{eq:H2} can then be  expressed as $3H^2(\tau)=8\pi G \rho(\tau)$, and corresponds to a global solution. Since $\rho(\tau)$ does not depend on $\chi$, any point can be chosen to be the centre $\vec{r}=0$.
In general, non-flat geometries or more sophisticated global topologies can be reproduced with a more general metric, that is, $C \neq 1$.
When considering spherical perturbations within a flat FLRW background, it is analytically appropriate to model the perturbation using a closed FLRW topology (i.e., with positive curvature, \( k = +1 \)) in GR. This approach accounts for the compensation of the surrounding background around the spherical perturbation (see Section\,\ref{sec:SC}). In particular, the positive local curvature in the FLRW metric reflects the curvature induced by the local overdensity. This helps maintain the isotropy and homogeneity of the perturbed solution, as it mirrors the curvature effects arising from the spherical symmetry of the perturbation.

\subsection{Uniform cloud}

The equations\,\eqref{eq:H2} and \eqref{eq:Mass} can also be used to solve non-homogeneous configurations. The simplest example is an expanding (or collapsing) uniform spherical mass distribution inside a comoving coordinate $\chi_*$, embedded in vacuum: 
\beq
\rho(\tau,\chi) = \left\{ \begin{array}{ll} 
\rho(\tau)  &  {\text{for}} ~ ~ \chi \leq \chi_*\\
0  & {\text{for}} ~ ~ \chi> \chi_* \\
\end{array} \right. \,.
\label{eq:rho1}
\eeq
Consequently, we have
\beq
M(\tau,\chi) = \left\{ \begin{array}{ll} 
\frac{4}{3} \pi r^3 \rho(\tau)  &  {\text{for}} ~ ~ \chi \leq \chi_*\\
M_{\rm T}  & {\text{for}} ~ ~ \chi> \chi_* \\
\end{array} \right. \,.
\label{eq:rho2}
\eeq
Here, we introduced
\beq
M_{\rm T} \equiv M(\chi=\chi_*)=\frac{4}{3}\pi  R^3 \rho(\tau) \,,
\label{eq:mT}
\eeq
where $R \equiv a\chi_*$ is the boundary or surface of the cloud.  
Inside $\chi<\chi_*$, this solution is identical to the infinite FLRW solution $r=a \chi$ with the same $\rho$. Any off-centre comoving observer inside $\chi_*$ still sees the same homogeneous and isotropic metric (\citealt{Gaztanaga:bhu1}).
The Hubble\textendash Lemaitre expansion law: $\dot{r} = H(\tau)r$ of equation\,\eqref{eq:H2} with $3H^2(\tau)=8\pi G \rho(\tau)$ is also the same in both cases within $\chi<\chi_*$.
The only difference is that for a mass distribution like equation\,\eqref{eq:rho1}, the FLRW solution only holds inside; outside $\chi_*$, there is empty space. Consequently, the total mass is finite $M_{\rm T}<\infty$.

The equivalence is given by {\it Birkhoff's Theorem} (see \citealt{BirkhoffH, faraoni}), which states that a sphere cut out from an infinite uniform distribution has the same spherical symmetry as the total distribution. Thus, the FLRW metric is both a solution to a global homogeneous (i.e. infinite $M_{\rm T}$) uniform background and also to the inside of a local (finite mass) uniform sphere centred around one particular point. The local solution is called the FLRW cloud (\citealt{Gaztanaga:bhu1}).

\subsection{Newtonian equivalence}
\label{sub:Newtonian_equivalence}

The total energy of a fluid with zero total energy $E=0$, uniform density and zero pressure in a Newtonian framework can be expressed as:

\bea
E=0= \frac{1}{2}\dot{\rt}^2 - \frac{GM(t,\rt)}{\rt} \quad \nonumber\\\Rightarrow \quad  \dot{\rt}^2(t) = \frac{8\pi G}{3} \rho_{\rm newt}(\tau)  \rt^2(\tau) \,,
\label{eq:E=0}
\eea
where $\rt$ is here the radius of the spherically symmetric mass distribution and $\rho_{\rm newt}$ is the usual Newtonian mass density. This equation\,\eqref{eq:E=0} is equivalent to equation\,\eqref{eq:H2} when replacing the surface of the relativistic cloud at $\rt \rightarrow r=R$.
Solving equation\,\eqref{eq:H2} for $R=R(\tau)$ requires two more equations. One is provided by the EoS $P=P(\rho)$, and the other is the continuity equation for $\rho$. For a perfect fluid, the latter can be derived from GR energy conservation $\nabla_\mu \boldsymbol{T}^{\mu\nu}=0$. 
The corresponding time component ($\mu=0$) for a uniform distribution gives:
\beq
\frac{\text{d}\rho}{\text{d}\tau} = \frac{\upartial \rho}{\upartial \tau} + 3~ H~(\rho+P) = 0 \,.
\label{eq:parta0}
\eeq
The radial equation ($\mu=1$) is:
\beq
\frac{\upartial p(\tau,\chi)}{\upartial \chi} = 0 \,,
\label{eq:pdot}
\eeq
which expresses the uniformity of the pressure in the comoving frame. We recall that we have chosen a comoving proper time $\tau$ so that $\boldsymbol{g}_{00}=-1$. The angular equations \(\mu=2,3\) are readily satisfied for a spherically symmetric fluid.

For vanishing pressure $P=0$,  the energy conservation equations \eqref{eq:parta0} and \eqref{eq:pdot} also apply to a uniform Newtonian cloud with \(M\) given by equation\,\eqref{eq:rho2} and replacing $r\rightarrow \rt$ and $\rho\rightarrow \rho_{\rm newt}$.
Thus, when written in the appropriate coordinates, the equations of motion describing the fully relativistic GR collapse of the flat FLRW cold cloud (defined by equations \ref{eq:FLRW}\textendash \ref{eq:rho1}) are equivalent to the equations of motion of the corresponding Newtonian problem with \(E=0\). This correspondence is well-known in the context of GR. Moreover, for the FLRW metric, this property does not depend on the particular choice of an EoS \citep[see][for a comprehensive description of the correspondence of GR and Newtonian solutions]{faraoni}. 

As is well known, speeds greater than $c$ are allowed in Newtonian mechanics and also appear in our simulations (see Section\,\ref{sec:simus}). For the description in the GR framework, velocities larger than $c$ pose no problem because they can be interpreted in a novel way. Since we are considering a background solution, the faster than light velocities are related coordinate velocities $\dot{r}$, which do not carry physical information. The Newtonian equivalence for $P=0$ is evident, and the interpretation is straightforward in GR. How to interpret different EoS with $P=P(\rho)\neq 0$ within the Newtonian treatment in the GR solution requires more work. In any case, the description of the GR solution with the use of a metric does not allow for information to travel faster than $c$.

\subsection{Spherical collapse}
\label{sec:SC}

\begin{figure}
\centerline{
\includegraphics[width=70mm]{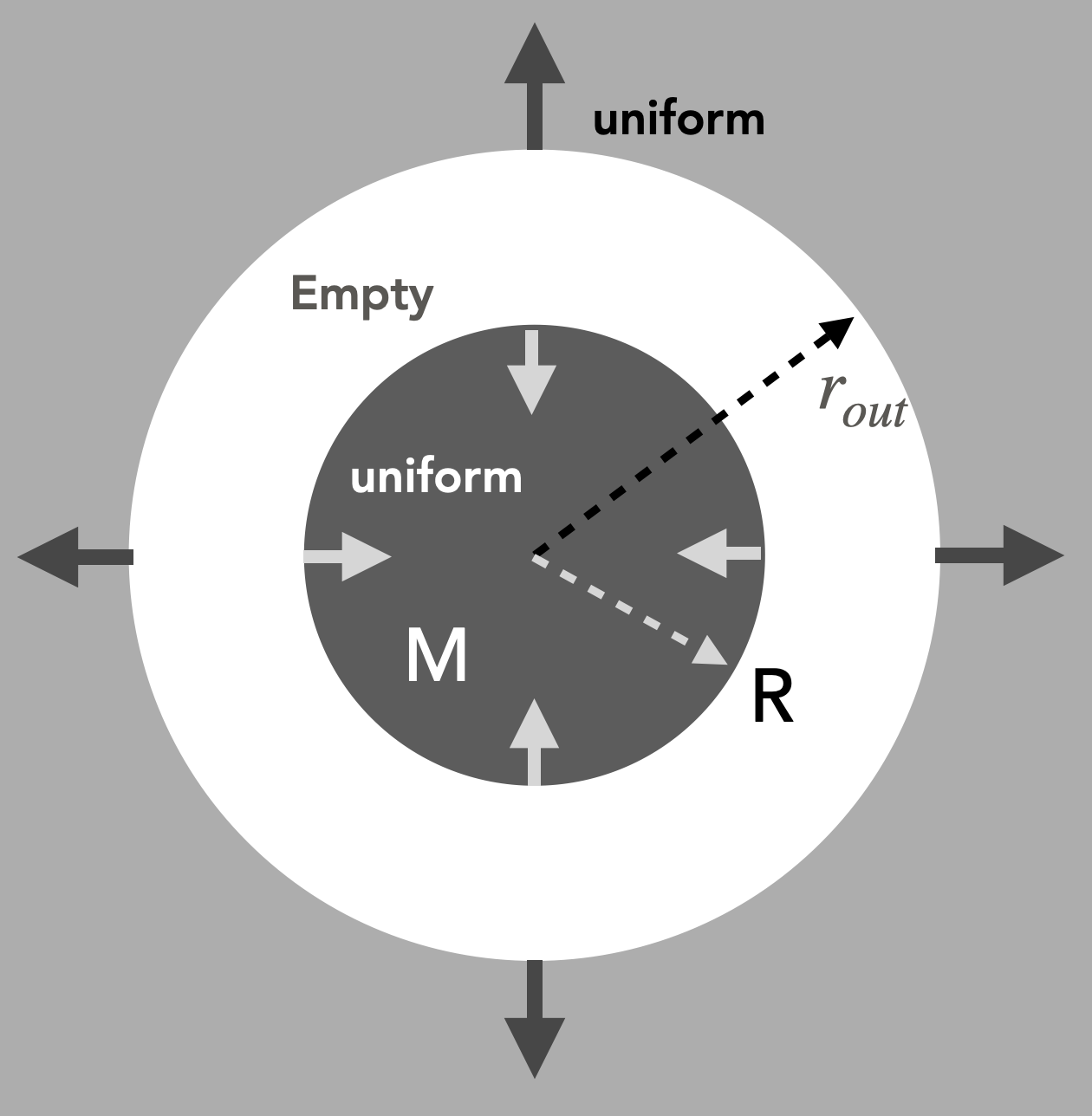}}
\caption{Graphical representation of the spherical collapse. There are three uniform spherically symmetric distributions: (i) outer region with radius larger than $r_\text{out}$ and mean density $\bar{\rho}$, (ii) inner region with radius less than $R$ and larger mean density $\rho=\bar{\rho}(1+\delta)$, and (iii) between $R<r<r_\text{out}$ empty space.}
\label{fig:collapse}      
\end{figure}

Consider a spherical perturbation of radius $R$ embedded in a background with mean density $\bar \rho$. The perturbed region has initiated their gravitational collapse and thus has a higher density $\rho=\bar\rho(1+\delta)$, which is initially only slightly higher than $\bar\rho$. We assume that the background with
mean density $\bar\rho$ follows the evolution of a flat universe dominated by matter (i.e., EdS universe).
To guarantee mass conservation,
the increase in density in $r<R$ has to be compensated for by an area of low or vanishing density. This low density zone (or void) is generated when the high density perturbation has initiated its collapse from its initial radius $r=r_\text{out}$ to its current value $r=R$ and left behind an empty space between $R<r<r_\text{out}$, see Fig.\,\ref{fig:collapse} for a graphical representation. The background
could be expanding or collapsing at a lower rate. In any case, mass conservation implies:

\beq
M = \frac{4\pi}{3} ~\left(r_\text{out}\right)^3~  \bar\rho ~= ~\frac{4\pi}{3}~R^3~\bar\rho( 1 + \delta)\,.
\label{eq:deltaM}
\eeq

Due to Gauss's theorem (or the corollary of Birkhoff's theorem) in GR, (see \citealt{BirkhoffH}), both spheres in Fig.\,\ref{fig:collapse} follow the FLRW* metric (where $*$ denotes the flat FLRW metric with the constraint $r = a \chi < R$), but each has a different expansion/collapse rate $H^2$. The outside has $k=0$, and the inside has $k>0$ ($k$ is known as the curvature: $k=1/\chi_{k}^2$, where $\chi_{k}$ is the comoving radius of curvature of the metric of equation\,\eqref{eq:FLRWk} below). 
The resulting exact dynamics of $R$ in GR also agree with Newton's law:
\beq
\ddot{R} = -\frac{GM}{R^2} = - \frac{4\pi}{3}  G \rho R =  -\frac{1}{2} \bar H^2  (1+\delta) R \,.
\label{eq:ddotR}
\eeq
In the last equality, we have used the Friedmann equation\,\eqref{eq:H2}, where $\bar H^2=\bar H_0^2 a^{-3}$ refers to the background expansion rate: $\dot{r}= \bar H r$ and $\ddot{r} = -\bar H^2r/2$, where $r \ge r_{out}$.
Combining with the second derivative of equation\,\eqref{eq:deltaM}:

\beq
\ddot{\delta} + 2\bar H \dot{\delta} - \frac{3}{2} ~\bar H^2~ \delta =  \frac{4}{3} \frac{\dot{\delta}^2}{1+\delta}
+ \frac{3}{2}~\bar H^2~ \delta^2 \,.
\label{eq:SCnl}
\eeq
This equation is exactly the same in Newtonian Cosmology (Newtonian dynamics in expanding coordinates) and in the GR dynamics, as given by the Raychaudhuri equation 
for an  irrotational and shear-free fluid (see \citealt{padmanabhan_2010}):
\beq
\nabla_\mu 
{\boldsymbol{q}^\mu} =
\frac{\text{d}\Theta}{\text{d}\tau} + \frac{1}{3} \Theta^2 = -\boldsymbol{R}_{\mu\nu} \boldsymbol{u}^\mu \boldsymbol{u}^\nu 
= -4\pi G(\rho+3P) + \Lambda \,,
 \label{eq:rPoisson} 
\eeq
where $\boldsymbol{q}^\mu$ is the acceleration four-vector, $\Lambda$ is the cosmological constant, $\Theta = \nabla_\nu \boldsymbol{u}^\nu$ and $\boldsymbol{u}^\nu$ is the  four-velocity.
The Raychaudhuri equation is the GR equivalent of the Newtonian Poisson equation. It is a purely geometric equation describing the evolution of the dilatation coefficient $\Theta$ of a bundle of nearby geodesics in proper time $\tau$. Together with the continuity equation (mass conservation), it results in equation\,\eqref{eq:SCnl}.
 In other words, {\em the Newtonian spherical collapse describes the actual GR dynamics when tidal effects are neglected}.

\subsection{Dynamics close to the ground state}
\label{sec:Ground state}

We assume here that the cold collapse to a classical singularity
$\rho \rightarrow \infty$ will be halted by the ground state of a matter degeneracy pressure, similar to what is expected from the Pauli exclusion principle in quantum mechanics. A ground state is characterized by a system dominated by its potential energy. 
In Appendix \ref{sec:Quintessence}, we show how for a generic system described by a single degree of freedom, which we represent by an effective scalar field $\psi$ in curved space\textendash time,  such a ground state corresponds to equation\,\eqref{eq:ground-state}.
In this case, we see from the equations\,\eqref{eq:relat1} and \eqref{eq:relat2} that $\rho = -P$ and the EoS of the fluid $P=P(\rho)=\omega \rho$ is characterized by $\omega =-1$.\footnote{Here, we assume the fluid to be barotropic, i.e. the pressure is assumed to not depend on the temperature. How far this holds during the collapse has to be investigated.} Combining equation\,\ref{eq:parta0} with 
equation\,\ref{eq:H2},
the acceleration in the FLRW metric is given by:
\beq
\frac{\ddot{a}}{a} = - \frac{4\pi G}{3} (\rho + 3P) \,,
\label{eq:ddota}
\eeq
which should be compared to equation\,\eqref{eq:rPoisson}.
We are interested in the collapse of cold, non-interacting matter characterized by initially vanishing pressure $P=0$. Once the collapse reaches a density at which matter cannot be compressed further, the pressure rises and causes the bounce. Just before the bounce, $\dot\psi<< V$, and from equations\,\eqref{eq:relat1} and \eqref{eq:relat2} we obtain $P=V(\psi)=-\rho$. This is also what happens during inflation or DE domination.

When $P$ changes from zero to $-\rho$, $\ddot{a}/a$ in equation\,\eqref{eq:ddota} changes from negative (accelerated collapse) to positive, describing a bounce and the subsequent expansion. Note that equation\,\eqref{eq:ddota} is independent of the curvature $k$. 

\subsection{A closed FLRW perturbation}

In the spherical collapse solution presented in Fig.\,\ref{fig:collapse}, we have assumed a closed local geometry for the perturbation with $k=1/\chi_{k}^2$. This is the standard approach in perturbation theory \citep[see e.g.][]{Bernardeau, padmanabhan_2010}. Such perturbation can be modelled as a closed FLRW metric:
\beq
\text{d}s^2=-\text{d}\tau^2 + a^2 \left(\frac{\text{d}\chi^2}{1-k\chi^2} + \chi^2 \text{d}\Omega^2\right) \,.
\label{eq:FLRWk}
\eeq
Note, that this metric is only valid for $\chi<\chi_{k}$, i.e. inside the closed perturbation. Outside, we assume that the metric is flat, as in the spherical collapse configuration of Fig.\,\ref{fig:collapse}. In proper coordinates, the curvature radius is $a\chi_{k}$ so that in the limit of large $a$, we map to the flat case (infinite curvature radius). The Friedman equation \eqref{eq:H2} for such GR perturbed space\textendash time is then given by:
\beq
H^2 + \frac{k}{a^2} = \frac{8\pi G}{3} \rho \,,
\label{eq:Friedmank}
\eeq
where the value of $k$ refers to the curvature scale at the time $\tau_0=0$. At this time, we choose $a=1$: i.e. $k=1/\chi_{k}^2$, where $\chi_{k}=\chi_*$ is the comoving curvature radius in equation\,\eqref{eq:rho1}.  
Since $H$ is related to the collapse/expansion velocity $\dot r$, and the latter should vanish right at the time of bounce, we also need $H=0$ at this time. Therefore, at the bounce we can define $\rho_{\rm B}$ as function of $a=a_\text{B}$:
\beq
\rho_{B} = \frac{3k}{8\pi G a_{\rm B}^2}\,.
\label{eq:ab}
\eeq
As explained in the subsection above, the conditions for the bounce are: change to positive acceleration, which requires $\omega<-1/3$ and is fulfilled by a ground state with $\omega=-1$ and zero velocity ($H=0$) at the bounce itself. When dominated by the ground state, $\omega=-1$, the expansion becomes exponential, but it turns into a regular expansion as soon as the density becomes smaller than the ground state $\rho_{B}$. 
In this work, we will study the Newtonian collapses of pressureless and polytropic fluids. 
In both the Newtonian and GR cases, we start from the same asymptotically free initial conditions. In the Newtonian framework, this corresponds to \( E = 0 \), while in GR, it corresponds to \( k = 0 \). The perturbation grows through gravitational instability, and we begin the numerical study at a point where the Newtonian solution already has some initial velocity, corresponding to a finite overdensity. At this stage, the GR equivalent problem has developed \( k > 0 \) due to the finite curvature radius associated with a finite perturbation mass.
Note that \cite{Gaztanaga:bhu1} focus on the junction conditions in equation\,\eqref{eq:rho1} for a flat $k = 0$ case (which is the right approximation for our expanding Universe or a flat expanding perturbation). The corresponding case with curvature during collapse was solved in §12.5.1 of \cite{padmanabhan_2010}.
We start from an initial perturbation with constant density up to $\chi=\chi_*$ in equation\,\eqref{eq:rho1}, which means that 
 $\chi_*=\chi_{k}$ in the spherical collapse model. The junction conditions in this case correspond to:
\beq
r_{\rm s} = 2G M_{\rm T} = (H_0/c)^2 \chi_{k}^3  \quad \Rightarrow \quad
k = [r_{\rm s}/c/H_0^2]^{-2/3} .
\label{eq:k}
\eeq
After the bounce, the curvature term is diluted by the exponential expansion and the 
expansion follows the flat case described above.

\subsection{White holes}
\label{sec:WH}
When the bounce happens inside the BH,
in both the Newtonian and GR solutions, the spherical perturbation expands and crosses back the gravitational radius \( r_{\rm s} \) from the inside out. This corresponds to a WH (white hole) solution. Although this behaviour is natural in the Newtonian framework, \cite{whiteholes} has argued that such a solution cannot exist in a GR scenario bouncing inside a BH. The standard argument supporting WH solutions is based on the time symmetry of GR equations: Since the metric is quadratic in time, one can reverse the arrow of time and convert a BH solution into a WH solution. Although this is theoretically true, we cannot arbitrarily change our choice of the time arrow as the solution evolves.

For an external observer outside the BH, the arrow of time points towards collapse. Once the bounce occurs inside the BH, the solution cannot transform into a WH because, according to the Schwarzschild metric, once the collapsing time arrow is fixed, nothing can escape the Schwarzschild radius \( r_{\rm s} \). The resolution to this conundrum, as detailed in \cite{whiteholes}, involves imposing a boundary condition on the GR action, given that the internal dynamics are confined within \( r_{\rm s} \). See Appendix \ref{app:boundary} for more details. This modifies Einstein's field equations of the expanding solutions from equation\,\eqref{eq:H2} to:
\begin{equation}
H^2 = \frac{2 G M(\tau,\chi)}{r^3} + \frac{1}{r_{\rm s}^2}.
\label{eq:H22}
\end{equation}
This boundary term is only needed for the  bouncing cloud that falls inside its gravitational radius $r_{\rm s}$. In our GR equivalent solution, we have ignored this term because it provides a very small contribution close to the bounce, but it is straightforward to include it if we want more accurate modelling after the bounce.
which prevents the expansion from crossing \( r_{\rm s} \).
This term corresponds to a reinterpretation of the measured \(\Lambda\) in cosmic expansion as a boundary term in the BHU solution: \( r_{\rm s} = \sqrt{3/\Lambda} \). According to this approach, a WH solution cannot emerge out of a BH. After the bounce, the expansion inside the BH is initially exponential, dominated by the ground state's negative pressure. It then transitions to being dominated by the more regular expanding density (which, in a more realistic scenario, will also contain radiation density) and eventually becomes dominated by the \(\Lambda\) boundary term, which is effectively determined by the initial mass \(M_{\rm T}\) of the collapsing perturbation: \( r_{\rm s} = 2GM_{\rm T} = \sqrt{3/\Lambda} \).
In the comoving frame, the expansion accelerates, but in the rest (de Sitter) frame, the expansion becomes asymptotically static within $r_{\rm s}$ as given by de Sitter metric (see \citealt{decceleration} for further details).

\section{Numerical simulations}\label{sec:simus}

As we have seen in the previous section, the solution for the collapse of an FLRW cloud is equivalent to solving the Newtonian problem outlined by equation\,\eqref{eq:E=0}, utilizing equation\,\eqref{eq:parta0} for vanishing pressure $P$. In this section, we make use of this equivalence and investigate the collapse numerically with the \textsc{CASTRO}\footnote{\textsc{CASTRO} is a Newtonian code that solves multicomponent compressible hydrodynamic equations for astrophysical flows using the unsplit second-order Godunov method. For details, see \citealt{Castro1}.} code.

In Section\,\ref{sub:pressureless_collapse}, we first concentrate on the pressureless collapse, for which we again highlight that the Newtonian formulation is not an approximation of the relativistic problem. Rather, both formulations are mathematically equivalent to each other. In GR, $r$ is interpreted as a surface coordinate $r=a(\tau) \chi$ rather than a physical distance as in the Newtonian framework. In the Newtonian solution, velocities $\dot{r}>1$ larger than the speed of light are allowed, and they indicate the formation of a BH. 
In the GR formulation, the coordinate distance $r$ is not related to the propagation of information. In contrast, the world lines of actual physical events or objects are described by equation\,\eqref{eq:FLRW} or the corresponding metric, and their propagation follows causality by definition. 
In this context, we note that, generally, no BH is formed in Newtonian simulations. However, we can estimate when a BH would form in GR as the time $t_\text{BH}$ when the matter of a given mass is located within its Schwarzschild radius $r_{\rm s}\equiv 2GM/c^2$ in the Newtonian solution. 

Once we have reproduced the pressureless case, we investigated the collapse of a polytropic fluid. Here, the pressure is negligible initially, and the analogy between Newtonian and GR solutions holds approximately. However, at some point, we postulate a strong interaction of quantum particles based on the Pauli principle, which leads to a significant increase of the pressure and finally to a bounce of the infalling matter. We then tentatively discuss the approximate Newtonian-GR equivalence for the second case in Section\,\ref{sec:Equiv2}.

\subsection{Pressureless collapse}
\label{sub:pressureless_collapse}

In the BHU interpretation, the observable Universe is expanding inside its own gravitational radius. This suggests that our observable Universe could originate from the collapse and subsequent bounce of a very large cloud. Given that the total mass $M_{\rm T}$ is extremely large (of the order of $M_{\rm T} \simeq 5 \times 10^{22}\, \rm M_\odot$), the density of the cloud at black hole formation  $\rho_{\rm BH}$ would be correspondingly very low:
\begin{align}
\rho_{\rm BH} &= \frac{M}{V} = \frac{3M}{4\pi r_{\rm s}^3} =\frac{3r_{\rm s}^{-2}}{8\pi G}
\simeq 3.9 \times 10^{-48}  \left[\frac{M}{M_{\rm T}}\right]^2 \, 
 \frac{\rm M_{\odot}}{\text{km}^{3}} \nonumber\\
\label{eq:BHrho}
\end{align}

For our Universe, this density is on the order of 1 atom per m$^3$ and motivates the idea that such an initial cloud existed within a very large manifold with comparable or even lower density. The original density must have been so low that the intermolecular separation was much larger than any potential source or interaction term, which implies that it was also extremely cold. In such a manifold \citep[which was called the Apollonian Universe by][]{Gaztanaga:bhu,Gaztanaga:bhu2}, small perturbations could give rise to a new type of primordial stars and black holes in hierarchical structures of different sizes. This perspective leads us to consider that the starting point of the Universe originated from such dilute, cold ($T = 0$ K) clumps of different sizes made of non-interacting, neutral hydrogen atoms. 

There is no specific reason to assume that this primordial cloud was composed of neutral hydrogen atoms, and other initial compositions are certainly possible. However, we select this scenario for concreteness, given that the results do not appear to depend significantly on the exact initial composition as long as the collapse remains cold. Although absolute zero temperature is not physically realizable, assuming a very low initial density makes this a reasonable approximation. Additionally, we assume that the cloud is spherically symmetric, homogeneous, non-rotating, and non-dissipative.

During the homologous collapse of this clump, the atoms feel each other only through their gravitational interaction, and if initially $T=0\,$K, no temperature will be created. As the collapse proceeds, the dust cloud reaches densities at which the wave functions of electrons in nearby atoms overlap. At this point, electrons become more and more degenerate as a result of the Pauli exclusion principle. Consequently, a pressure component is built up and ionizes the atoms. This phenomenon is completely temperature-independent and is called pressure ionization \citep{kothari_pressure_ionization}. This electron degeneracy pressure is responsible for the stability of White Dwarfs and allows hydrostatic equilibrium in stellar cores until they reach masses greater than $\approx1.44\,$M$_\odot$, the famously-known Chandrasekhar mass limit \citep{Chandrasekhar}. For more massive, spherically symmetric mass distributions, the electron degeneracy pressure is insufficient to prevent gravitational collapse. In our simulations of collapsing clumps starting at masses of at least $5\,$M$_\odot$, which should ensure that electron degeneracy pressure does not affect the collapse significantly. With masses significantly above the Chandrasekhar limit, our configurations are too massive to be influenced. Additionally, we start the collapse at very low densities ($E=0$), and once the densities reach values where electron degeneracy becomes important, the atoms have already acquired a significant infall velocity. Therefore, the collapse cannot be stopped by the electron degeneracy pressure. 

Since it is not physically realizable, the particular case of a pressureless fluid is generally not included in hydrodynamical codes and causes numerical problems. However, practically a very small pressure does not affect the evolution of the collapse of a dust cloud. We follow \citealt{mnm} and use a gamma-law EoS with an initial pressure 
\beq 
P_0 << \frac{4\pi G}{\gamma}\rho_{\rm 0}^2 r_{\rm 0}^2, 
\eeq
where $\gamma$ is the relation of specific heats of the gas, and $\rho_{\rm 0}$ and $r_{\rm 0}$ are the initial, uniform density and radius of the cloud, respectively.
In order to save numerical resources, we do not start the simulations right at the onset of the collapse. We rather approximate the initial phase of the collapse, where pressure can be neglected, with the free fall solution for $P=0$. The velocity at any point of the evolution is given by $u_0 = c\sqrt{r_{\rm s}/r_{\rm 0}}$. The collapse time of the analytical solution is
\beq \label{eq:tc}
\tau_{\rm c} = \frac{2\sqrt{(r_{\rm 0}^3/r_{\rm s})}}{3c}
\eeq
and will serve as the reference time for our numerical simulations. 

We set up our initial data inside a simulation box of radius $r_{\rm b}$ and place a uniform-density ($\rho_{\rm 0}$) sphere of radius $r_{\rm 0} < r_{\rm b}=300\times10^6\,$cm in the centre. The exterior of the sphere is filled with a very low-density medium ($\rho_\text{amb}$). In between the two regions, the transition profile is defined as \cite{Castro1}:
\begin{align} 
    \label{(4.4)}
    \rho_\text{newt} = \rho_{\rm 0} - \frac{\rho_{\rm 0}-\rho_\text{amb}}{2}\left[ 1 + \tanh{\left( \frac{r-r_{\rm 0}}{h} \right)} \right]\,,
\end{align}
where $h$ is the smoothing length, which we fix to be $h=4\times 10^4$ cm in our simulations.

\begin{figure}
    \centering
    \begin{subfigure}{\columnwidth}
        \includegraphics[width=\columnwidth]{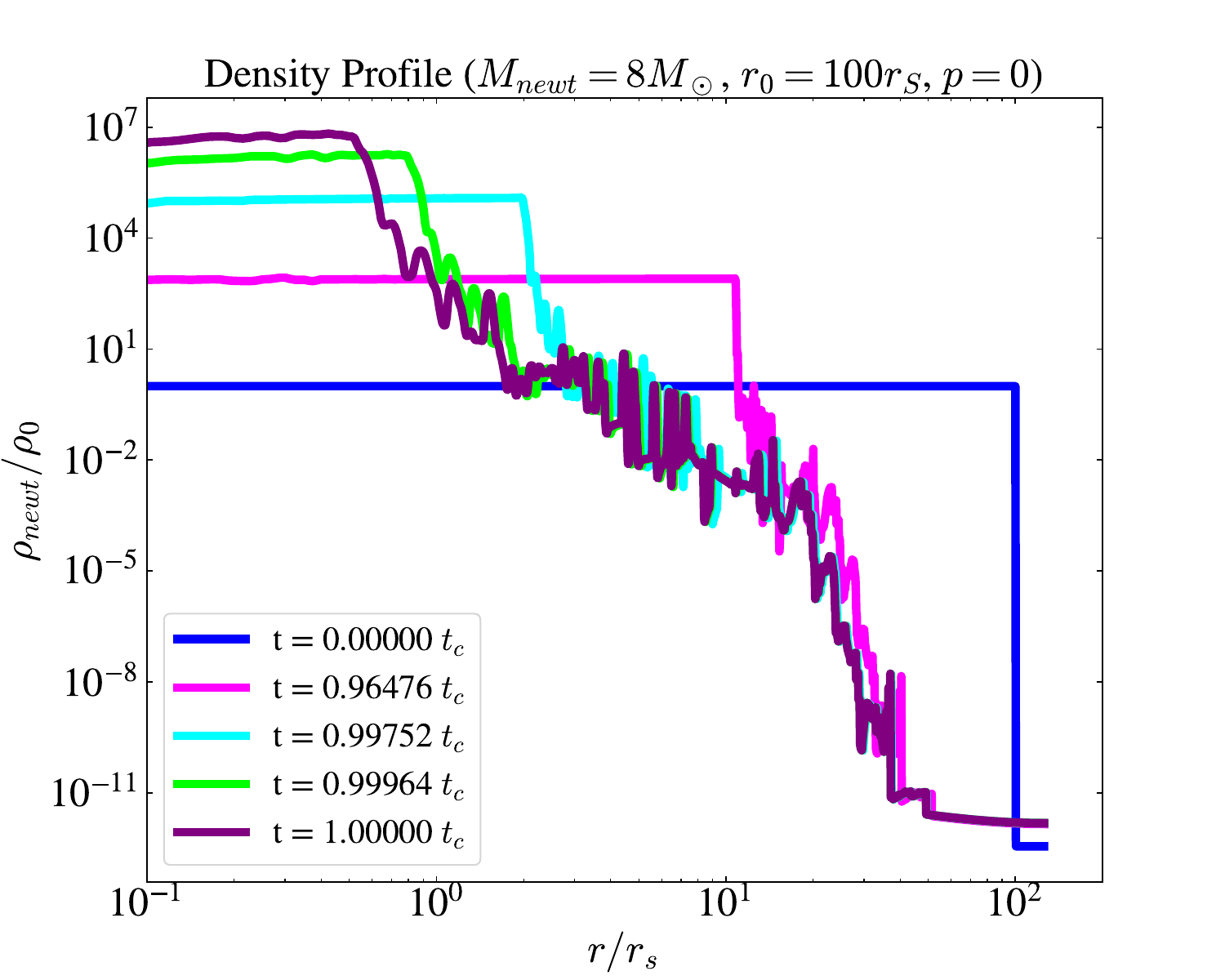}
    \end{subfigure}
    \begin{subfigure}{\columnwidth}
        \includegraphics[width=\columnwidth]{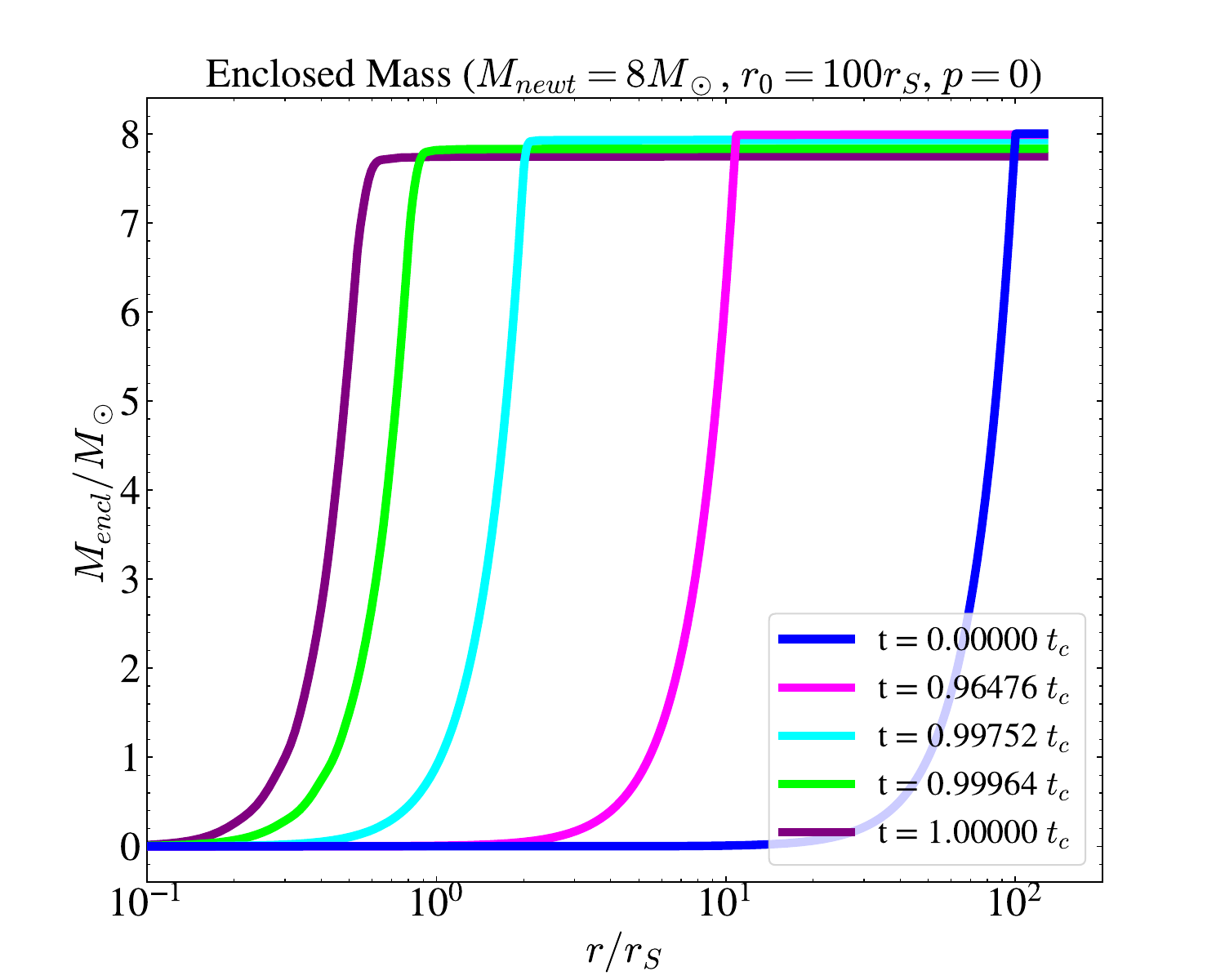}
    \end{subfigure}
    \begin{subfigure}{\columnwidth}
        \includegraphics[width=\columnwidth]{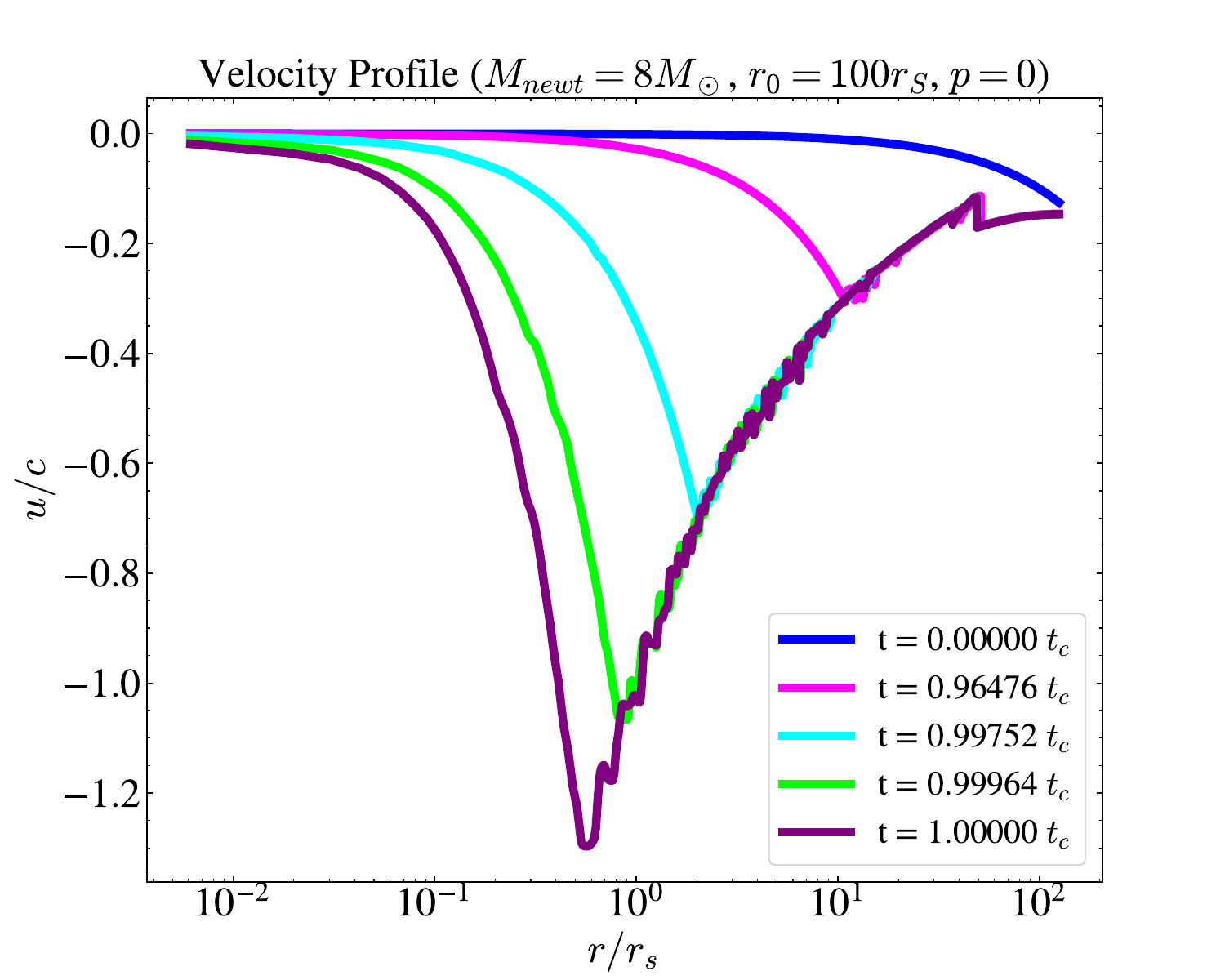}
    \end{subfigure}
    \caption{Snap shots of the radial distributions of $\rho_\text{newt}$ (top panel), enclosed mass (central panel) and velocity (bottom panel) of the simulations of the collapse of the pressureless dust cloud of $8\,\rm M_\odot$ starting at $r=100r_{\rm s}$.}
    \label{fig:collapse1}
\end{figure}

We simulate the cold or pressureless collapse of an $8\,$M$_\odot$, spherically symmetric, and homogeneous dust cloud starting at $r = 100r_{\rm s}$. We obtain the corresponding values of initial density and collapse time at that radius from the theoretical solution for a pressureless collapse. The corresponding values are given in the first row of values of Table\,\ref{tab:bounce2}. We set the initial pressure to $P_0 = 10^{10}\ $dyne cm$^{-2} << \frac{4\pi G}{\gamma}\rho_{\rm 0}^2 r_{\rm 0}^2=2.26\times10^{27}$ dyne cm$^{-2}$. 
We stop the simulation at the estimated collapse time $t_{\rm c} = 0.0532\,$s. The simulation uses a Courant-Friedrichs-Lewy (CFL) number of $0.5$, and the radial resolution is set to 10240 grid points. At the outer boundary, we use outflow boundary conditions, while at the centre, symmetry conditions are applied. The gravitational potential is calculated using monopole gravity $\phi_{\rm G}(R)=-GM/R$, which represents the spherically symmetric component of the gravitational field.

\subsubsection*{Results}

In the top panel of Fig.\,\ref{fig:collapse1}, we plot the density as a function of the radius $r$. The initial configuration (blue curve, $t=0.0 t_{\rm c}$) is a solid sphere with $r=100r_{\rm s}$. The sphere has a sharp transition to the low-density exterior. During evolution, the density increases and the radius decreases as expected.
The density within the sphere remains roughly constant until the mass is concentrated within $r=r_{\rm s}$. At this point, the resolution in the centre limits the accuracy of our simulation. In addition, unavoidable numerical diffusion leads to a smearing of the initially sharp edge between the high-density sphere and the ambient matter. However, the edge is still qualitatively well defined and visible at all times. The density jump at the edge decreases from $12$ orders of magnitude in the beginning to $3$ orders at $t=t_{\rm c}$. Note that the exterior also evolves with time. 

In the last time step $t=t_{\rm c}$, the densities reached supranuclear densities $\rho_\text{newt}\approx10^7\rho_{\rm 0}=2.8\times10^{15} g/cm^3$.
As the simulation proceeds, the total mass of the system is concentrated in smaller and smaller radii (central panel of Fig.\,\ref{fig:collapse1}). Once the radius of the sphere is comparable to $r_{\rm s}$, the limited numerical resolution causes a loss of some of the mass. We could increase the resolution or use adaptive methods. However, our interest here is the time up to the collapse, which is reproduced well. The formation of a black hole is indicated by a free-fall velocity $|u|>c$. As we can see in the bottom panel of Fig.\,\ref{fig:collapse1}, this is reached around $t=0.99964 t_{\rm c}$ (green curve). 

\begin{figure}
    \centering
    \begin{subfigure}{\columnwidth}
        \includegraphics[width=\columnwidth]{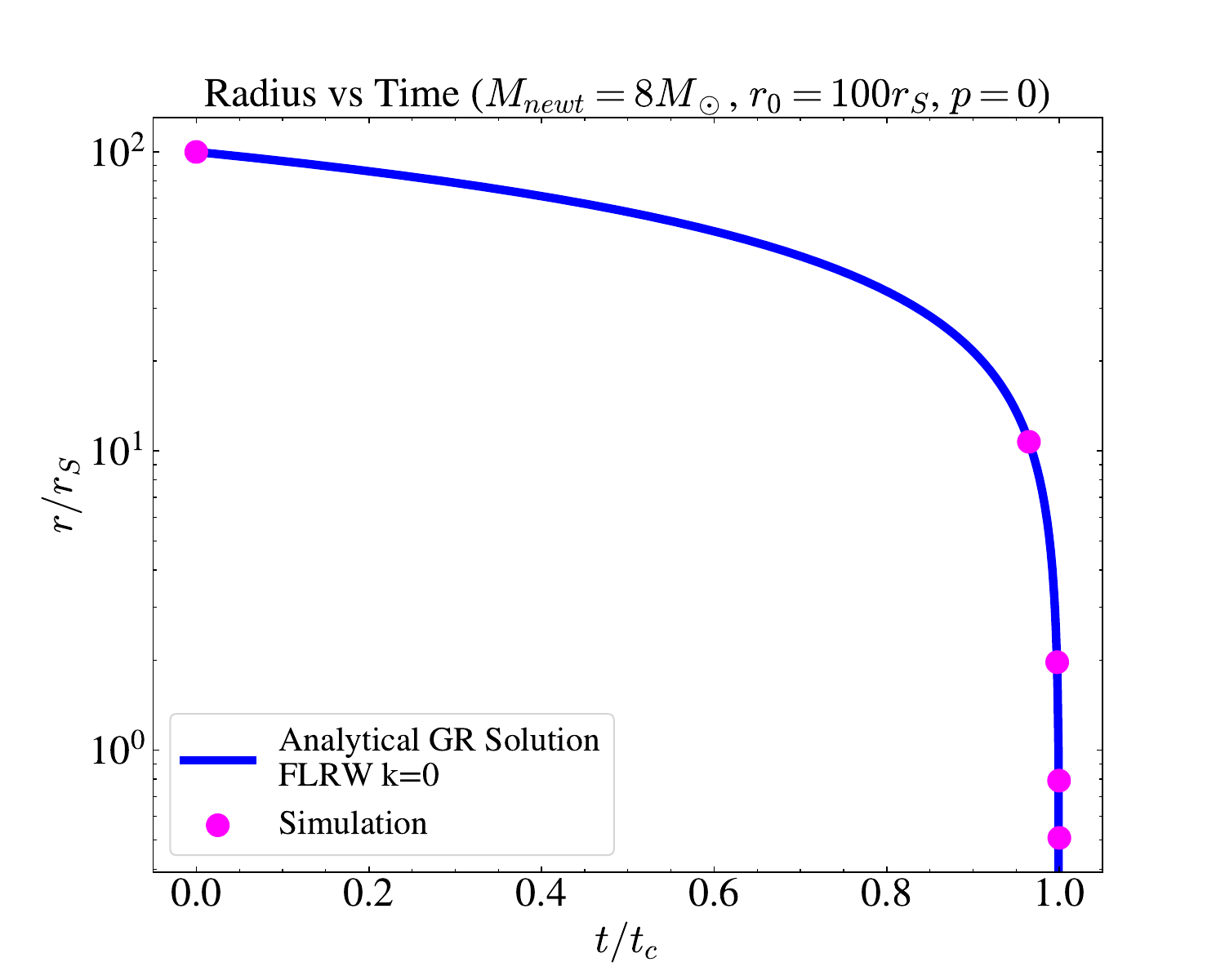}
    \end{subfigure}
    \begin{subfigure}{\columnwidth}
        \includegraphics[width=\columnwidth]{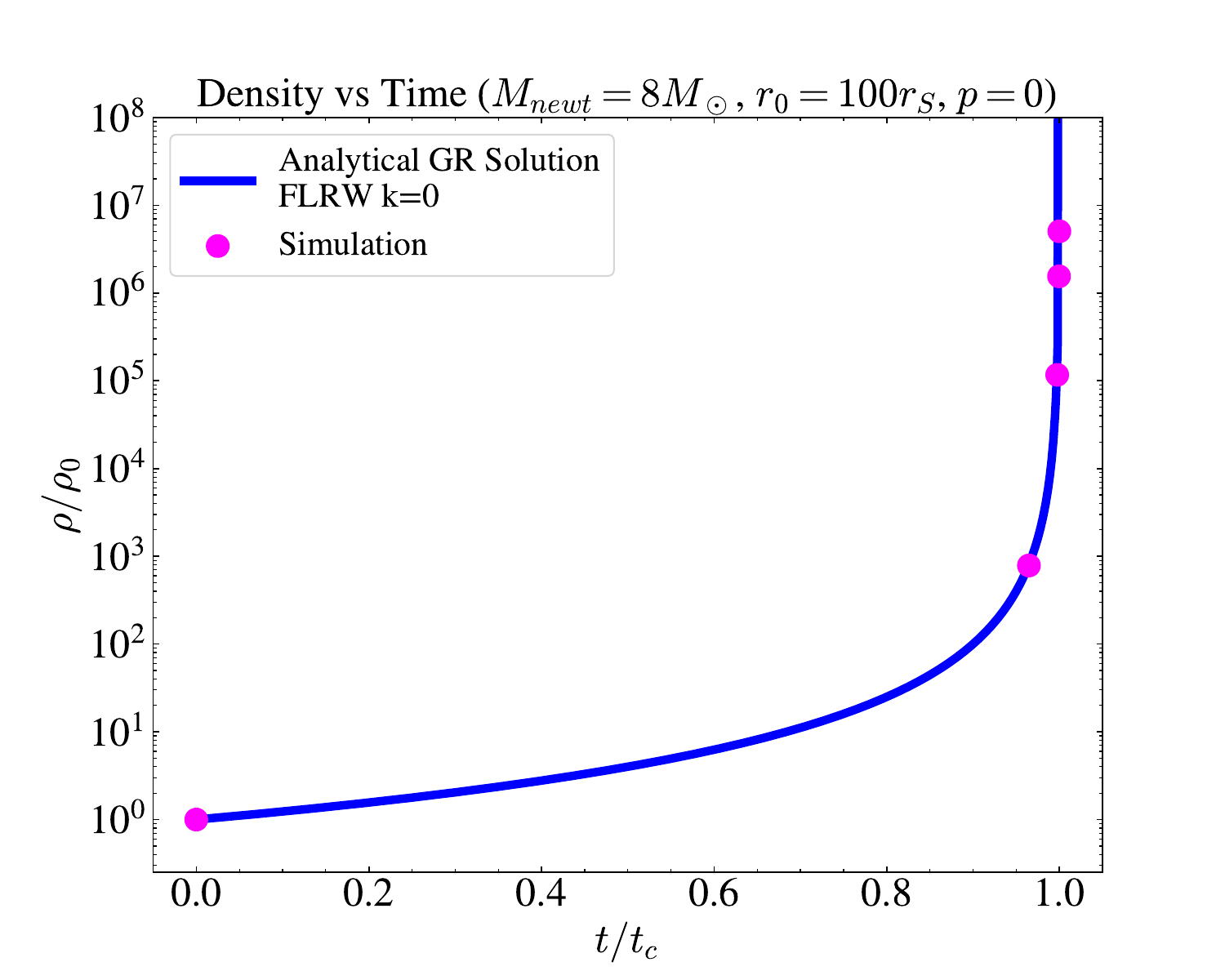}
    \end{subfigure}
    
    \caption{The radius and density of the collapsing cloud as functions of time in the top and bottom panels, respectively. The magenta dots represent the simulation results, and the blue lines represent the exact GR solutions.}
    \label{fig:collapse2}
\end{figure}

We can compare our simulations with the theoretical results predicted by GR. For this purpose, we plot the radius and the corresponding density of the collapsing spherical cloud as functions of time in Fig.\,\ref{fig:collapse2}. From the Friedmann equation \eqref{eq:E=0}, we get the radius $r(\tau)$:
\begin{align}
    \label{(4.16)}
    r = r_{\rm 0}\left(1-\frac{\tau}{\tau_{\rm c}}\right)^{2/3} = \left(\frac{3}{2}(\tau_{\rm c}-\tau)\right)^{2/3}.
\end{align}
In our simulations, the radius is defined as the location of the edge of the mass density profile, and the density of the sphere is approximated as the average of all densities inside the sphere enclosed by this radius. The comparison of the simulation (magenta dots) with the analytical prediction (blue line) for the radius (density) is depicted in the top (bottom) panel of Fig.\,\ref{fig:collapse2}. We obtain a good agreement up to short before $t=t_{\rm c}$. In the theoretical GR solution, $r$ tends to zero, and the density $\rho_R$ increases towards infinity. We anticipate that in the presence of non-zero pressure, the GR solution will not be exact anymore, and the collapse time will be longer. This is already reproduced in our numerical results, which include a small but unavoidable pressure. The results of the simulations lag behind the theoretical curve after $t>0.999 t_{\rm c}$.

\begin{figure}
\includegraphics[clip,width=\columnwidth]{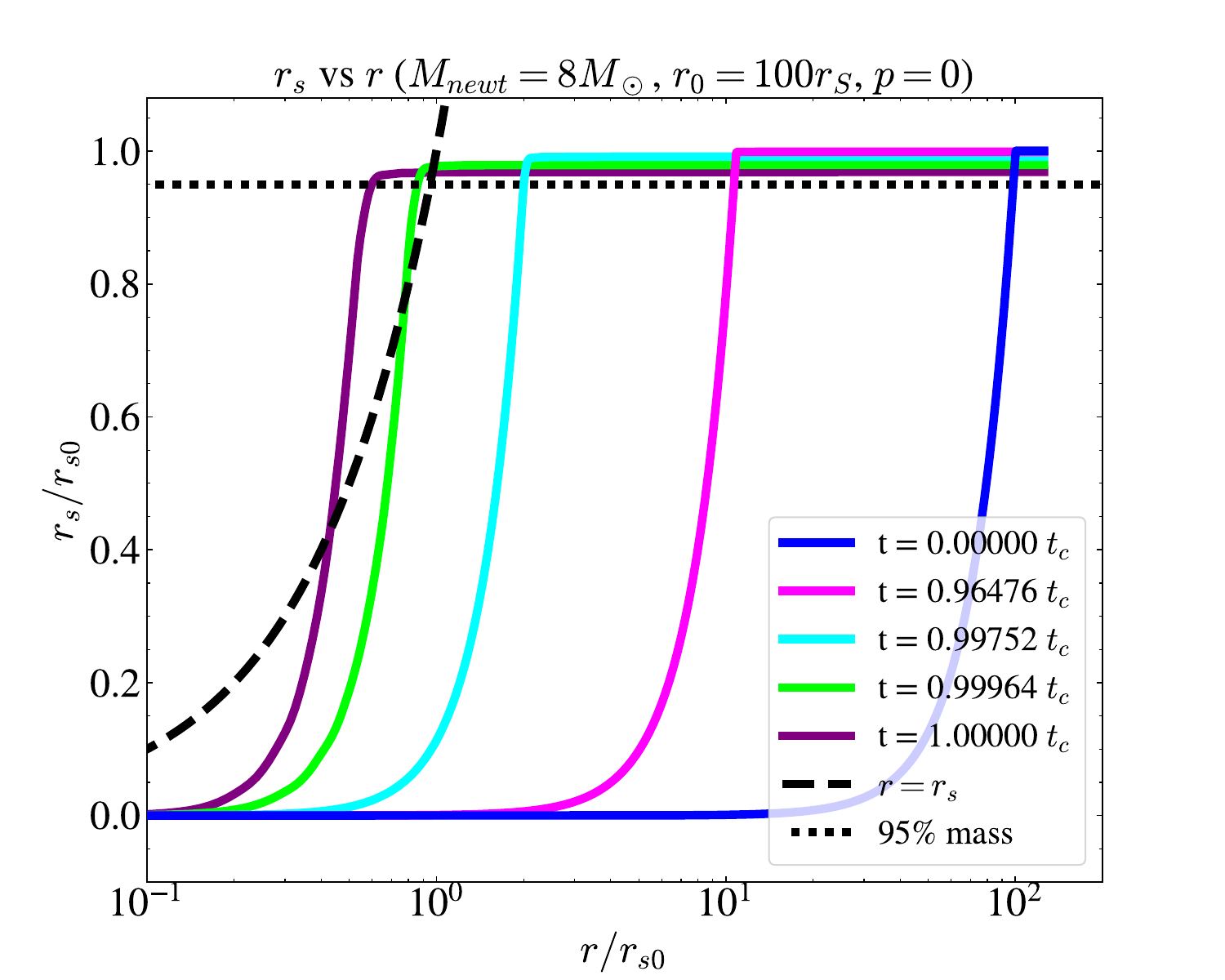} 
\caption{Rescaled Schwarzschild radius, $r_{\rm s} / r_{\rm s0}$ as a function of the rescaled radial coordinates $r / r_{\rm s0}$. Note that $r_{\rm s0}$ is the Schwarzschild radius of the initial mass, whereas $r_{\rm s}$ is the Schwarzschild radius for a sphere of radius $r$ with an enclosed mass $M_
\text{encl}(r)$. The black dashed line indicates the location where $r=r_{\rm s}$ and the dotted line represents the $r_{\rm s}$ of $95$ per cent of the initial mass.
}
\label{fig:collapse3}
\end{figure}

We have seen that the numerical resolution in our simulations affects the total mass inside $r$, and the edge of the sphere becomes less sharp with time. In order to take this into account, when considering the formation of BHs in the GR analogy, we take a more conservative approach. Instead of checking whether the entire mass is within its Schwarzschild radius, we check for each radius whether the mass $M_\text{encl}(r)= 4\pi \int_0^r  \rho(\rt) (\rt)^2 \text{d}\rt$ enclosed by a given radius $r$ has reached its Schwarzschild radius $r_{\rm s}\left(M_\text{encl}\right)=r_{\rm s}(r)$ or whether it is still more extended $r>r_{\rm s}\left(r\right)$. In Fig.\,\ref{fig:collapse3}, we plot the Schwarzschild radius of the mass distribution with $8\,$M$_\odot$ at each radius at different times, both rescaled by the Schwarzschild radius of the entire mass $r_{\rm s0}$. Initially (blue line), the radii are much larger than the corresponding $r_{\rm s}$. With time, the collapse proceeds, and the radii approach the Schwarzschild radii. Once the coloured curves, representing the evolution, cross the black dashed line where $r=r_{\rm s}$ a BH would form in the GR case. This has happened for the last two steps shown, and the system's total mass is located inside its Schwarzschild radius. As we will see later in our simulations, some of the models bounce back before $r=r_{\rm s}$ is reached. We find that a threshold value of $95$ per cent of the mass is a good indicator to see whether the collapse of the mass distribution has reached $r=r_{\rm s}$ or not.

\begin{table*}
    \begin{tabular}{ c c c c c c c c c c} \hline  
         $M$& EoS& $\gamma$& $K$& $r/r_{\rm s}$&  $\rho_{\rm 0}$& $\rho_\text{amb}$&  $r_{\rm 0}$&  $r_{\rm b}$&   $\boldsymbol{u}_0$ \\  

         (M$_\odot$)& & & & & ($10^{12}$g cm$^{-3}$)& (g cm$^{-3}$)& ($10^6$ cm)& ($10^6$ cm)& ($10^{10}$ cm s$^{-1}$) \\ \hline 
         \multirow{2}{*}{${8}$}& Approximately&\multirow{2}{*}{-}&\multirow{2}{*}{-}& \multirow{2}{*}{ $100$}&  \multirow{2}{*}{$2.8\times10^{-4}$}&\multirow{2}{*}{ $10^{-4}$}&\multirow{2}{*}{  $239$}& \multirow{2}{*}{$300$}&\multirow{2}{*}{ $-0.299$} \\
         &zero pressure\\ \hline 
          \multirow{3}{*}{$5$}& \multirow{3}{*}{Polytropic} &$2$&$10^5\,$cm$^2$&  \multirow{3}{*}{$2$}&   \multirow{3}{*}{$89.6$}&  \multirow{3}{*}{$10^{-4}$}& \multirow{3}{*}{$2.99$}&  
          \multirow{3}{*}{$3.2$}&
          \multirow{3}{*}{$-2.11$} \\ %
          &&$2.5$&$10^{-2}\,$cm$^3$\\
          &&$3$&$10^{-9}\,$cm$^4$\\ \hline
           \multirow{3}{*}{$20$}& \multirow{3}{*}{Polytropic} &$2$&$10^5\,$cm$^2$&  \multirow{3}{*}{$2$}&   \multirow{3}{*}{$5.6$}&  \multirow{3}{*}{$10^{-2}$}& \multirow{3}{*}{$11.9$}&  
           \multirow{3}{*}{$13$} &
           \multirow{3}{*}{$-2.11$} \\ 
          &&$2.5$&$10^{-2}\,$cm$^3$\\
          &&$3$&$10^{-9}\,$cm$^4$\\ \hline
           \multirow{3}{*}{$1000$}& \multirow{3}{*}{Polytropic} &$2$&$10^5\,$cm$^2$&  \multirow{3}{*}{$0.05$}&   \multirow{3}{*}{$143$}&  \multirow{3}{*}{$10^{-1}$}& \multirow{3}{*}{$14.9$}& \multirow{3}{*}{$18$}&  \multirow{3}{*}{$-13.4$} \\ 
          &&$2.5$&$10^{-2}\,$cm$^3$\\
          &&$3$&$10^{-9}\,$cm$^4$\\ \hline
    \end{tabular}
\caption{Initial conditions and employed EoS for the different models simulated.}
\label{tab:bounce2}    
\end{table*}

\subsection{Collapse with a polytropic EoS}
Even for a zero-temperature gas, the negligible pressure approximation applied in the previous section breaks down at some point. First, electrons become degenerate and create temperature-independent pressure. By performing simulations with a zero-temperature-white-dwarf EoS, we found that the collapse for sufficiently high masses ($M\ge 5$ M$_\odot$), is basically indistinguishable from the zero pressure case.
However, already at nuclear saturation densities, nucleon degeneracy can halt the collapse and initiate a bounce for masses up to several $10's$ of M$_\odot$. This scenario is actually observed at the end of the life of massive stars as supernova explosions. Analogously, we assume that for more massive objects, a yet unknown pressure, based on the Pauli principle for quantum objects at the subnuclear scale, prevents further collapse and leads to a bounce.
To simulate this effect, we investigate the behaviour of collapsing clouds with a polytropic EoS. For a perfect fluid, the latter is given by:
\beq
P = K \rho^\gamma \,,
\label{eq:poly}
\eeq
where $K$ is the polytropic constant and $n=1/(\gamma-1)$ is the polytropic index. This kind of EoS can be a good approximation for nuclear degenerate matter with $\gamma=2\dots 3$ (see \cite{Lattimer_Prakash_eos}). Though not taking all subtleties of the nucleonic interaction properly into account, the polytropic EOS has the advantage of being simple. Since we also do not know what happens at much higher densities than nuclear saturation density, we can encode all our ignorance on what may happen at supranuclear densities into the parameters of a polytropic EoS. We do not aim to explain the realistic bounce of a collapsing cloud of matter under these conditions. However, the polytropic EoS allows us to study qualitatively under which conditions a bounce may happen and what the properties of this bounce are.

For the $\gamma$ values in the range $2\dots3$, we choose the $K$ parameter of the polytropic EoS such that the pressure resembles a strong repulsion once nuclear densities are reached. The combinations of $\gamma$ and $K$ are given in the third and fourth columns of Table\,\ref{tab:bounce2}. The EoS are constructed such that they have the same pressure at $\rho=10^{14}\,$g cm$^{-3}$, which is slightly lower than the nuclear saturation density $\rho_{\rm ns}$.  This choice ensures that the evolution until reaching densities of the order of $\rho_{\rm ns}$ is not affected significantly by the pressure and that all simulations are comparable until these densities. The larger $\gamma$, the stiffer the EoS and the stronger the pressure increases with density. So, for larger $\gamma$, we would expect a bounce to occur at lower densities. Similarly, bounces at higher (lower) pressures for a given density can be achieved by choosing a lower (higher) value of $K$. 

For the study of the effect of the different polytropic EoS, we consider three different masses $M=5$, $20$ and $1000$ M$_\odot$. The other initial conditions of the mass distributions are chosen similar to the ones used in Section \ref{sub:pressureless_collapse}. We place a barotropic fluid with given $\gamma$ and $K$ of uniform density $\rho_{\rm 0}$ in a sphere of radius $r_{\rm 0}$. This sphere is placed inside a simulation box of radius $r_{\rm b}>1.1\times r_{\rm 0}$ filled with a very low density, $\rho_\text{amb}$. The temperature is $T = 0K$, and the initial pressure is obtained from the polytropic EoS. The initial quantities are given in Table\,\ref{tab:bounce2}. 

To compare our results to the previously obtained ones, we use the time of the collapse (equation\,\ref{eq:tc}) of a pressureless cloud as a reference. However, now we do not stop the simulations at $\tau_{\rm c}$ and continue the simulation beyond it. For the polytropic EoS, the simulations inevitably create a bounce. The pressure increases with density to arbitrarily high values, which does not allow for further compactification.  Consequently, at some point, the pressure obtained during the collapse of any mass configuration will reach such high values that it can balance the gravitation and lead to a bounce. 
Once the bounce occurs, we may encounter a number of numerical problems, like strong oscillations of the central density or the bounced mass starting to leave our predefined outer grid boundary $r_{\rm b}$. In both cases, we stop the analysis, even if the simulation continues. 
The time-stepping is controlled by the CFL condition with a CFL number of $0.5$, and the spatial resolution is set to 10240 grid points. As before, symmetry conditions are used at the centre, and outflow boundary conditions are set at the outer boundary of our grid. 

Note that collapses of polytropic mass distributions have already been studied by \cite{bondi}. They find that for $5/3>\gamma>4/3$, the system can either oscillate and be in stable equilibrium or lead to a bounce. Whereas, the configuration with $\gamma=5/3$ invariably leads to a bounce in the Newtonian case. However, the initial conditions studied by \cite{bondi} are different from the ones we use here. In our case, the pressure of the whole cloud is determined by the polytropic EoS, whereas Bondi uses a polytropic EoS only for the centre of the cloud. Additionally, the equations used in \cite{bondi} work only outside the gravitational radius. However, our results are in line with the ones found by Bondi, supporting the close correspondence between Newtonian and GR results.

\subsubsection*{Results}

\begin{figure}
    \centering
        \includegraphics[width=\columnwidth]{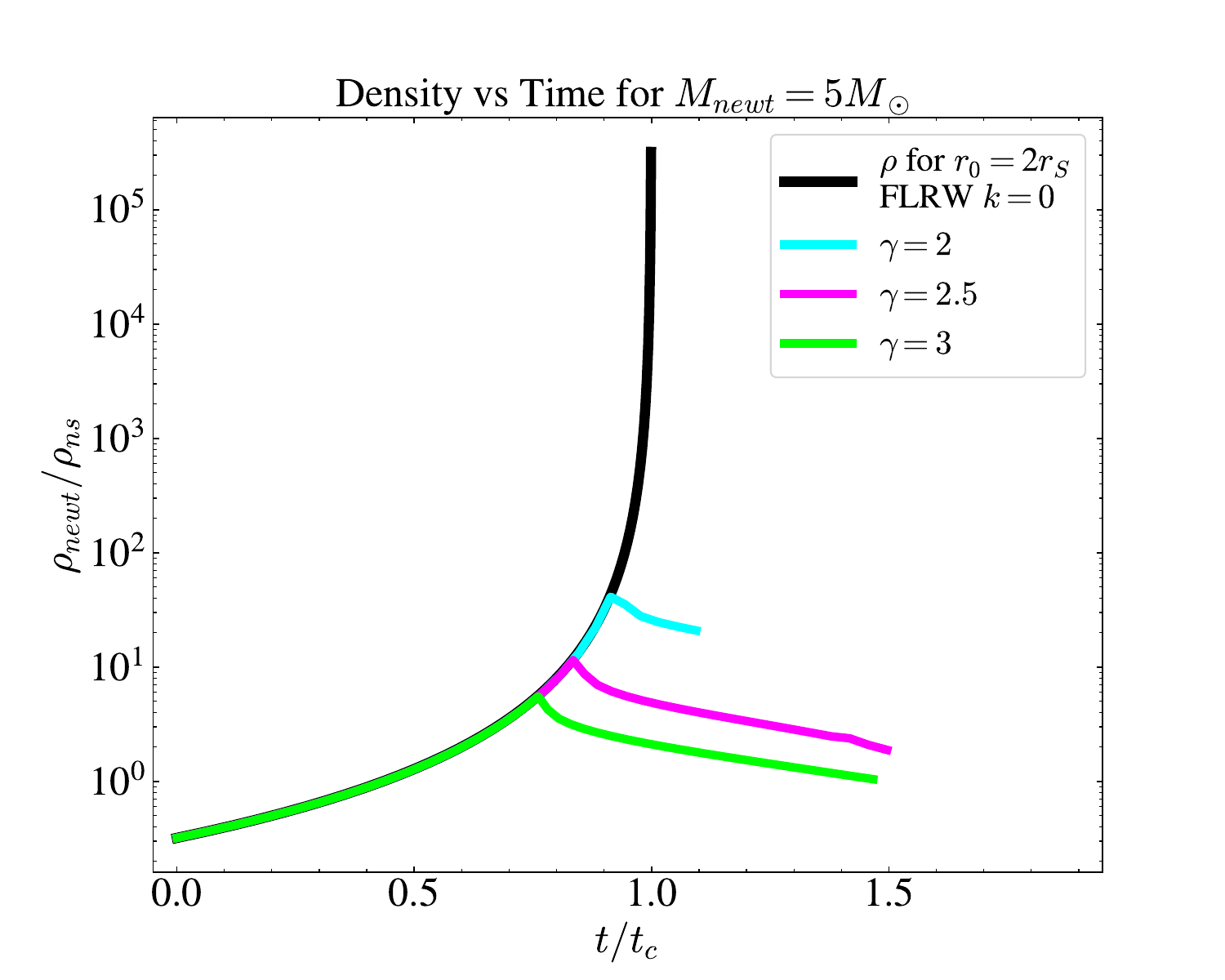}
        \includegraphics[width=\columnwidth]{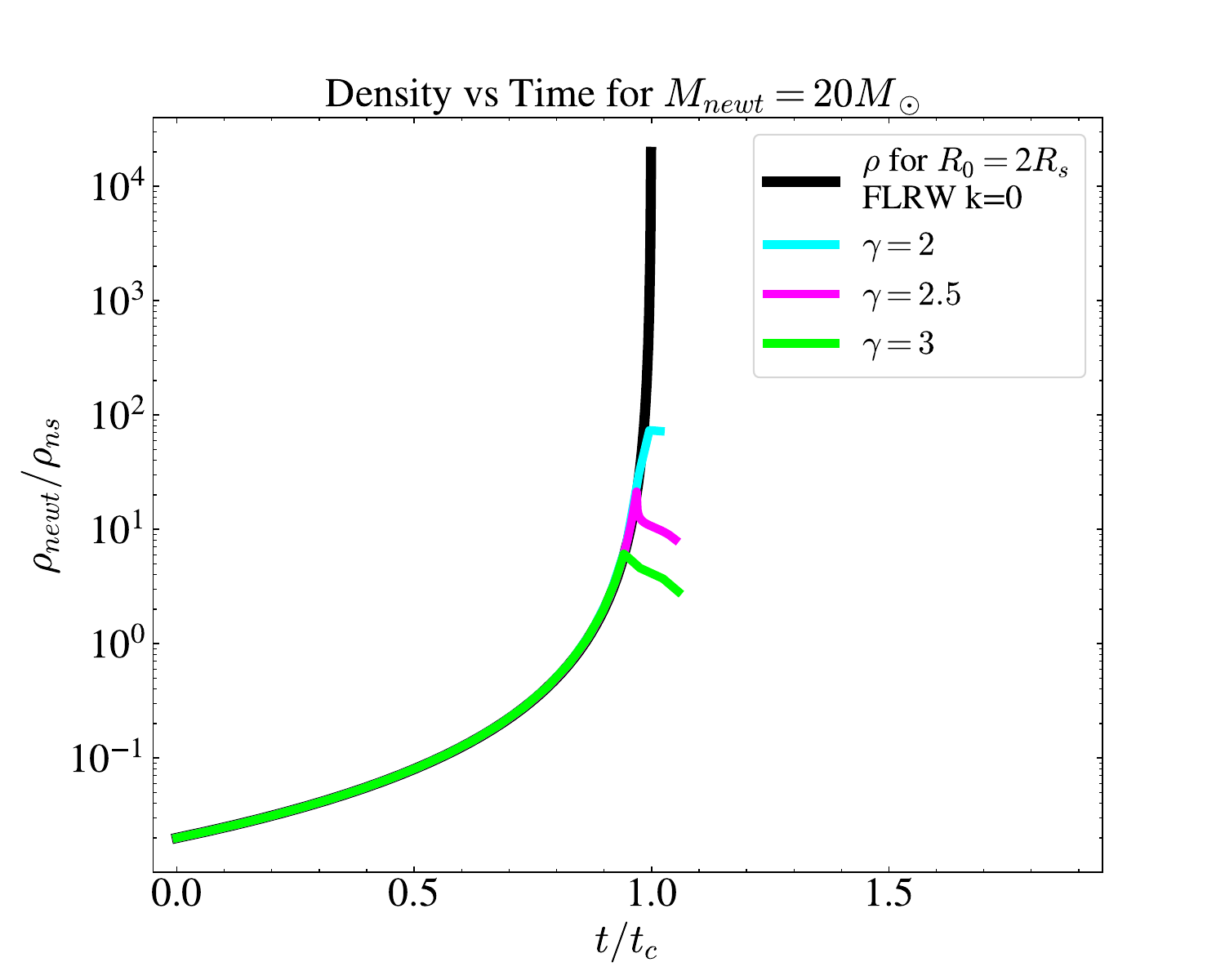}
        \includegraphics[width=\columnwidth]{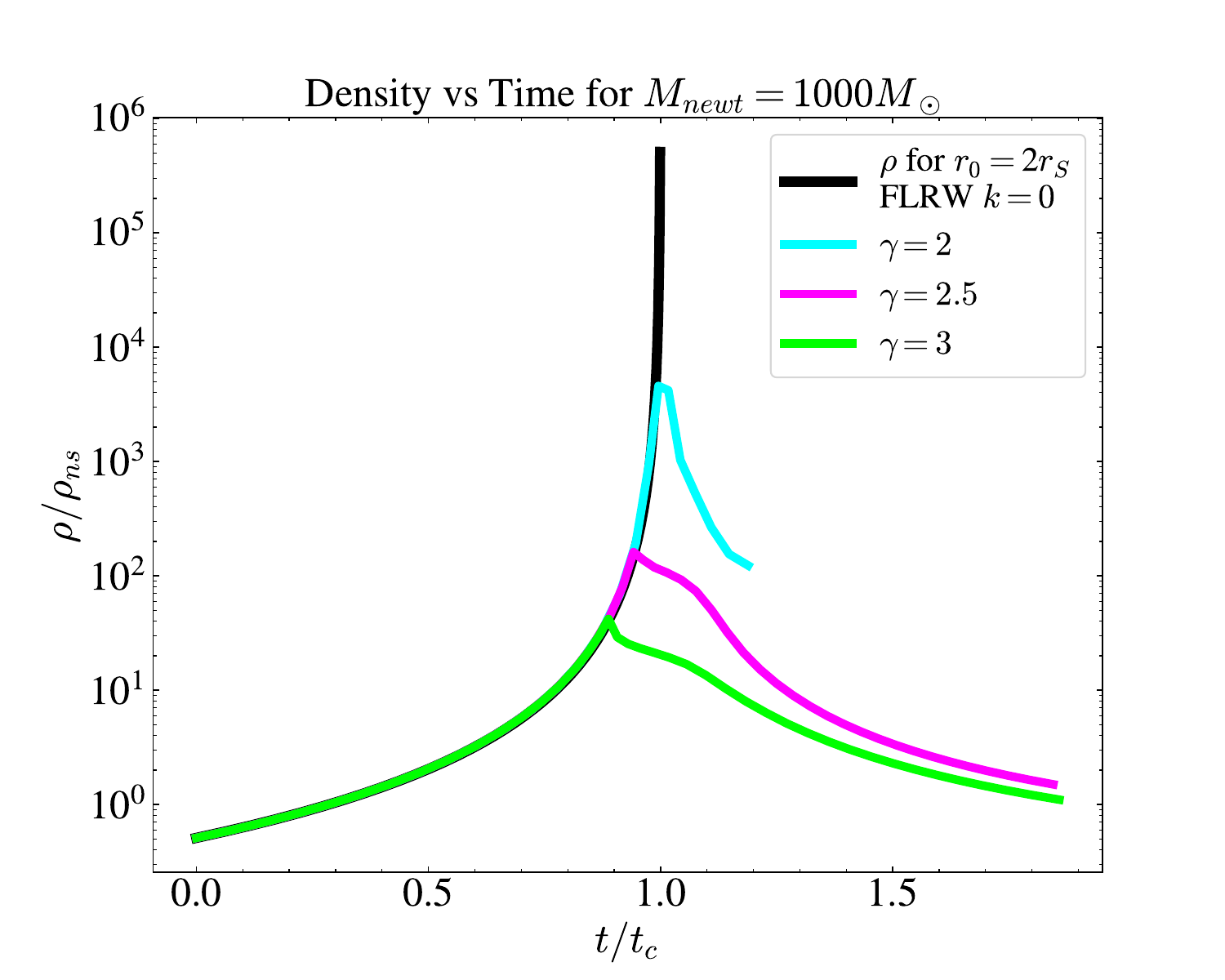}
    \caption{Time evolution of the central density for the $5$, $20$, and $1000\,$M$_\odot$. The simulations started at $r=2r_{\rm s}$ for the first two clouds (top and middle panels) and at $r=0.05r_{\rm s}$ for $M=1000\,$M$_\odot$ (bottom panel). The solid black line represents the analytic results for pressureless collapses. 
    }
    \label{fig:bounce3}
\end{figure}

We performed a total of nine simulations consisting of models with three different masses and each with three different polytropic indices (see Table\,\ref{tab:bounce2}). In Fig.\,\ref{fig:bounce3}, we plot the central densities as a function of time. The black line is the density prediction of the zero-pressure case. As we can see, the curves of all nine configurations follow the black line until after we reach densities $\rho_\text{newt}>\rho_\text{ns}$. Initially, the pressure from the polytropic EoS is still very low, and the evolution is as if there was no pressure.
Note that in the figure for $1000$ M$_\odot$, we start the simulation for radii $r<r_{\rm s}$, such that in GR, the simulations would have started within a BH from the beginning. 
In all cases, the central densities rise to values higher than $\rho_{\rm ns}$, and then, sooner or later, all models bounce back. This bounce can be recognized as a drop in density. Note that an apparent higher density in our polytropic simulations, compared to the pressureless case, is simply caused by a limited number of the time snapshots we used for drawing the coloured curves, while the black line is analytic.  

\begin{figure}
    \centering
    \includegraphics[width=\columnwidth] {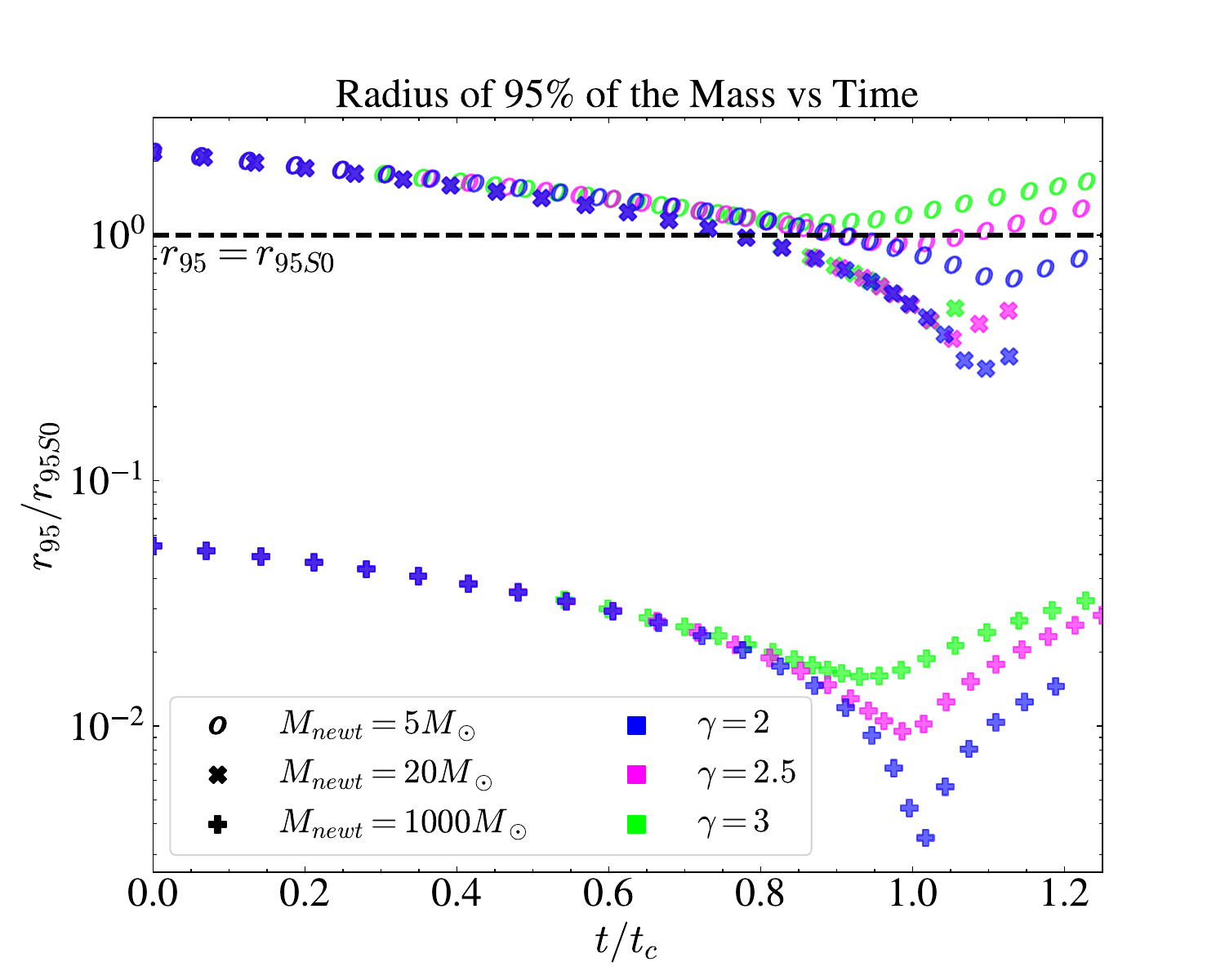}
    \caption{Radius enclosing $95$ per cent of the total mass as a function of time. $r_{95S0}$ is the Schwarzschild radius for the $95$ per cent of the total mass. Note that the collapse time $t_{\rm c} = \tau_{\rm c}$ is different for different masses.}
    \label{fig:bounce6}
\end{figure}

Similarly to what we have discussed before for numerical viscosity, the pressure introduces a complication when determining the radius of the initially sharp edge of the mass distribution. With increasing pressure, the outermost matter falls in slower than in the pressureless case, and the edge smears out. 
Therefore, when checking whether a potential BH may form in a GR solution for these configurations, we define $r_{\rm 95}$ as the radius enclosing $95$ per cent of the total mass and plot its ratio with the Schwarzschild radius of this mass $r_{\rm 95}/r_{95S0}$ as a function of time in Fig.\,\ref{fig:bounce6}. When $r_{\rm 95}/r_{95S0}=1$, we expect that a BH forms in GR.
In all models, the radius starts to expand after an initial phase with decreasing $r_{\rm 95}$. We call the time when this happens bounce time $t_{\rm B}$.
For almost all models, the bounce occurs within the Schwarzschild radius. Only the model with $M=5\,$M$_\odot$, and $\gamma=3$ bounces before reaching $r_{\rm s}$. The bounce happens so early for this model because the pressure grows the fastest for $\gamma$ = 3, and, in the case of a $M=5\,$M$_\odot$ sphere, it can counteract the gravitational pull before total mass falls into its gravitational radius $r_{\rm s}$.

\setlength{\tabcolsep}{3pt}
\begin{table}
    \centering
    \begin{tabular}{|c|c|c|c|c|}
        \hline
         M  & Quantity & $\gamma = 2$ & $\gamma = 2.5$ & $\gamma =3$ \\
        \hline
         & $t_{\rm B}/t_{\rm c1}$ & $0.914$ & $0.835$ & $0.761$ \\
        $5\,$M$_\odot$ & $\rho/\rho_{\rm ns}$ & $41$ & $12$ & $6$ \\
        & $P$ [$10^{36}\,$dyne cm$^{-2}$] & $13$ & $4$ & $2$ \\ \hline
        & $t_{\rm B}/t_{\rm c2}$ & $0.996$ & $0.969$ & $0.942$ \\
        $20\,$M$_\odot$ & $\rho/\rho_{\rm ns}$ & $74$ & $21$ & $6$ \\
        & $P$ [$10^{36}\,$ dyne cm$^{-2}$] & $43$ & $19$ & $3$ \\ \hline
        & $t_{\rm B}/t_{\rm c3}$ & $0.996$ & $0.942$ & $0.888$ \\ 
        $1000\,$M$_\odot$ & $\rho/\rho_{\rm ns}$& $4580$ & $162$ & $42$ \\
        & $P$ [$10^{38}\,$dyne cm$^{-2}$] & $1650$ & $29$ & $8$ \\ \hline
    \end{tabular}
    \caption{Bounce time $t_{\rm B}$, central density $\rho$, and central pressure $P$ at bounce for different masses and $\gamma$. $t_{\rm c1}$, $t_{\rm c2}$ and $t_{\rm c3}$ are the collapse time-scales for the $5$, $20$ and $1000\,\rm M_\odot$ configurations, respectively.}
    \label{tab:bounce3}
\end{table}

The obtained parameters of the $t_{\rm B}$, density, and pressure at bounce for all models are given in Table \ref{tab:bounce3}. The densities (and pressures) increase with mass and decrease with increasing $\gamma$. Consequently, the configuration with $M=1000\,$M$_\odot$ and $\gamma=2$ reaches the highest densities and pressures. We expect this trend to hold for larger masses. When extrapolating to a configuration with a mass as large as that of the entire Universe, we expect to reach densities many orders of magnitude higher than the nuclear saturation density at bounce. This is far beyond any reachable energies we can study experimentally today. Further work in high-energy physics is needed to develop an EoS to describe the matter at those densities. 

Furthermore, higher pressures and densities are reached for a given mass for a lower $\gamma$ value or a higher polytropic index. This is intuitive since a higher $\gamma$ means a higher stiffness. The cloud cannot be deformed as much as in softer EOS, with lower $\gamma$. Equally, the bounce time for a given mass strongly depends on the polytropic index. It takes longer for the bounce to occur for a lower polytropic index since the pressure increase is slower.

As we have seen in this section, in the Newtonian problem, there is no problem for the matter to expand past its Schwarzschild radius after it had collapsed to within $r<r_{\rm s}$ (see Fig.\,\ref{fig:bounce6}).

\section{Discussion}
\subsection{Newtonian versus GR equivalence}
\label{sec:Equiv2}

As discussed in Section \ref{sub:Newtonian_equivalence}, the equations of motion describing a GR collapse of a flat FLRW cloud have mathematically equivalent Newtonian equations for a system with zero total energy. In the absence of pressure, the GR solution can be satisfactorily reproduced by Newtonian simulations, as we showed in Section\,\ref{sub:pressureless_collapse}. When the pressure does not vanish, the situation is more complex. The interpretation of the Newtonian solution in terms of the GR equivalent is not straightforward and requires further work. 

In this work, we focus on the dynamics of spherical collapse in \S\ref{sec:SC}, where tidal effects are absent. As argued from equation\,\eqref{eq:SCnl}, the Newtonian spherical collapse accurately captures the full GR dynamics for a perfect fluid with an arbitrary EoS as long as the pressure and density are homogeneous. In GR, our aim is to simulate a flat ($k=0$) background, which corresponds to $E=0$ in the Newtonian solution. However, an important distinction arises once we introduce a perturbation. A spherical GR perturbation on a flat background is characterized by a positive local curvature $k=1/(a\chi_{k})^2$ in the equation\,\eqref{eq:FLRWk}. In the corresponding perturbed Newtonian solution, $k\neq0$ would introduce a constant energy term $E=r_{\rm 0}/\chi_{k}$ (the relation between the initial radius $r_{\rm 0}$ and the initial curvature radius $\chi_{k}$ at $a=1$). This would pose a problem if we want to compare the Newtonian to a flat background GR solution because, in Newtonian physics, energy conservation is a fundamental principle. In contrast, in GR, there is no direct counterpart. To ensure that the Newtonian simulation aligns with the flat-background GR space asymptotically, we retain the $E=0$ initial condition. This distinction becomes vital when deriving the equivalent GR EoS from a given Newtonian EoS, as the curvature and energy terms are handled differently between the two frameworks.

Given the equivalence between the GR and Newtonian equations, under certain approximations, we can estimate what to expect in GR from the results of the equivalent Newtonian simulation. This kind of mapping allows us to leverage the simplicity of Newtonian simulations while gaining insight into the corresponding GR behaviour, providing a useful approximation for more complex relativistic scenarios.

Let us now estimate the qualitative behaviour expected for the GR solution from our Newtonian simulations with a non-zero-pressure EoS. As a first step, we use Newtonian simulations with a polytropic EoS to estimate the scale factor $a(t)$. Using equation\,\eqref{eq:r=achi} and the initial condition $a(0) = 1$, we have $a(t) = R(t) / r_{\rm 0}$. With $a(t)$, we can estimate the Hubble rate using equation\,\eqref{eq:H2}: $H=\dot{a}(t)/a(t)$.
Once $H$ is determined, we can calculate the relativistic density $\rho(t)$ using the Friedmann equation \eqref{eq:Friedmank}. Finally, the pressure $P(t)$ can be obtained from the continuity equation \eqref{eq:parta0}. With $P$ and $\rho$ given, the relativistic EoS is determined. 

\begin{figure}
    \centering
\includegraphics[width=\columnwidth] {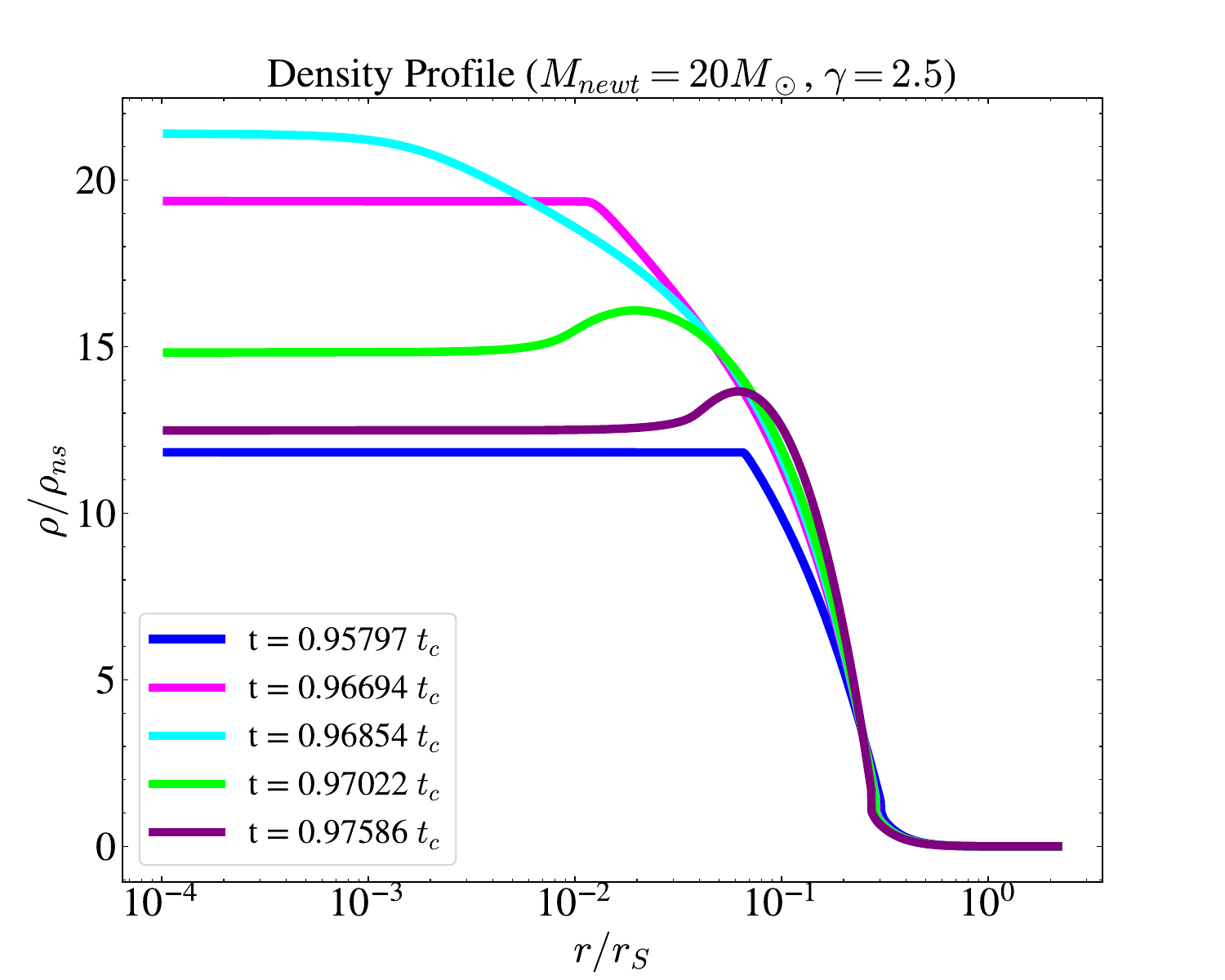}
    \caption{Density profile for the model with $M=20\,\rm M_\odot$ and $\gamma=2.5$ before and during the bounce.}
    \label{fig:bounce8}
\end{figure}

We discuss our findings as exemplary for the model $M=20$ M$_\odot$ and $\gamma=2.5$. Before discussing the results, we note that finding the exact radius $R(t)$ where the collapsing shell terminates at a given time is tricky once $P \ne 0$. The reason is that once there is pressure, the density does not remain constant within the cloud, and the matter at the outer edge of the shell collapses somewhat slower than in the centre. This is depicted in Fig.\,\ref{fig:bounce8}, where we show the density as a function of radius for a few time-steps. There is still a region of almost constant density in the very centre of the cloud. However, this constant density region no longer accounts for all the matter of the cloud. With subsequent collapse, the mass having constant density decreases further and, at the time of bounce, the density decreases monotonically from the inside out (cyan curve in Fig.\,\ref{fig:bounce8}). Actually, this moment of maximum central density is only the onset of the bounce, and most of the matter is still in-falling. The process takes time, and a hump in the density moves outwards and marks the region where bounced matter encounters the still in-falling material. Note that since the density is not constant, the assumption of an FLRW cloud no longer holds, and our results and interpretations are only approximations.

Without a well-defined edge of the density profile, we estimate the equivalent radius of our collapsing non-zero-pressure cloud by using four-thirds of the mass-weighted radius
\beq
R_\text{equiv} = \frac{4}{3} \frac{\int_0^{r_{\rm b}} (4\pi \rho r) r^2  dr }{M} \,.
\label{eq:R_equiv}
\eeq
With this $R_\text{equiv}$, we can calculate the scale factor and the Hubble rate. For our $20$ M$_\odot$ model equation\,\eqref{eq:k} gives $k = 6.29\times 10^6$ s$^{-2}$. This $k$ corresponds to a curvature radius of $\chi_{k} \simeq 1.19 \times 10^{7}$ cm, which is lower than the Hubble Horizon $c/H_0$ at the initial condition: $H_0=H(t=0) \simeq 1.69 \times 10^7$ cm. This means that, initially, the cloud is still inside its Hubble horizon, and everything is causally connected.


\begin{figure}
    \centering
        \includegraphics[width=\columnwidth]{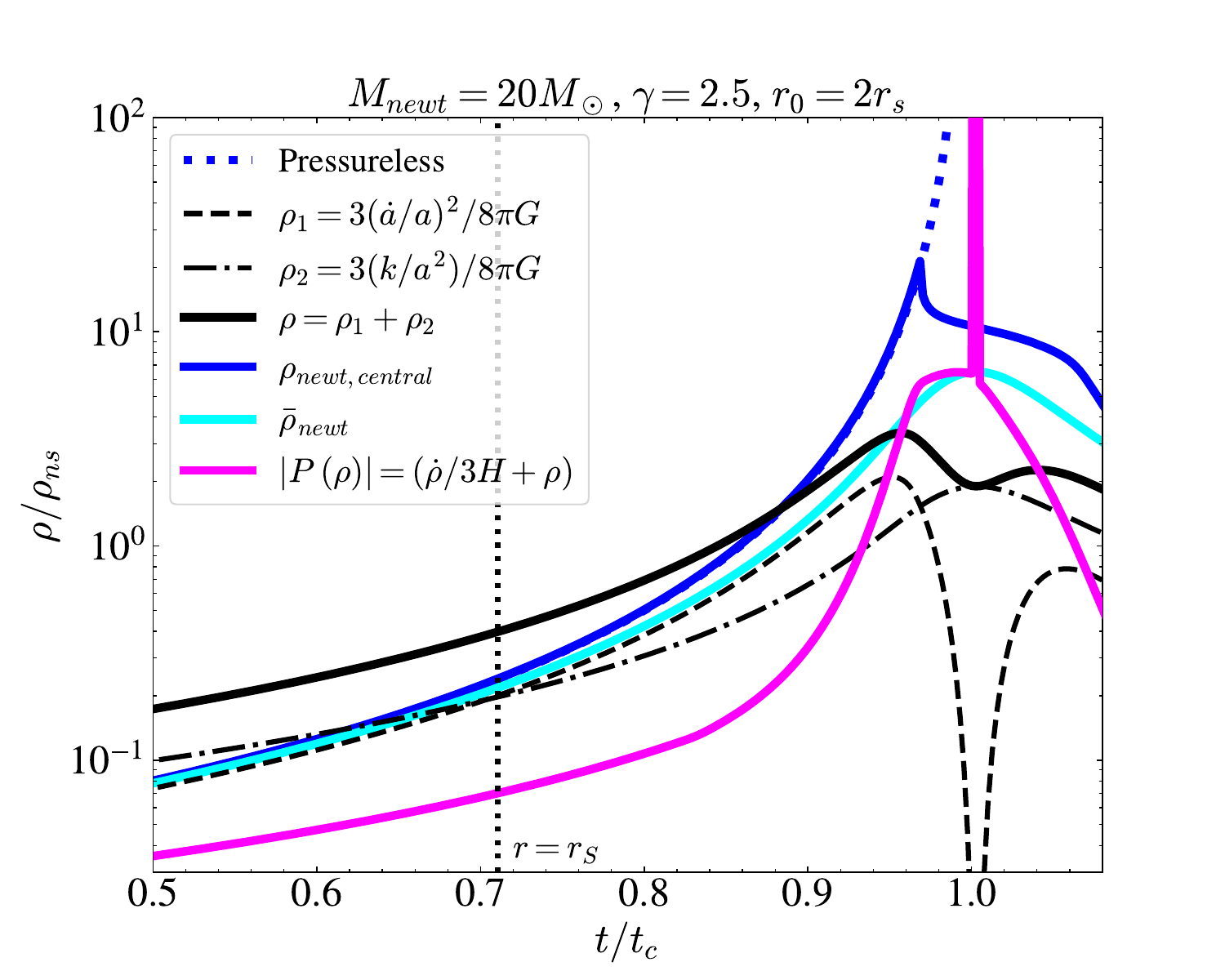}
    \caption{Different densities as a function of time for the collapse of a cold cloud with an initial mass $M=20\,\rm M_\odot$ and polytropic EoS with $\gamma=2.5$. The exact pressureless solution, the central density, and the average density of the Newtonian simulation are given as the dotted blue, solid blue and solid cyan curves, respectively. The black dotted vertical line marks the time when the BH forms.
    The approximate equivalent relativistic (FRLW) case (solid black curve) has two contributions: $\rho_1$ (dashed black) from the kinetic energy term $H$ and $\rho_2$ (dot\textendash dashed black) from the curvature term $k$. For comparison, we also plot the absolute value of the equivalent relativistic pressure $P$ (magenta curve), which diverges when $H=0$.}
    \label{fig:bounce9}
\end{figure}

To estimate the behaviour of the bounce in GR, we plot different representative densities in Fig.\,\ref{fig:bounce9}. The dotted blue line represents the exact FLRW solution of the pressureless collapse, and the solid blue line represents the corresponding central density of the Newtonian simulation with the polytropic EoS. As expected, the pressure has a vanishing influence for $t\ll t_{\rm c}$. However, once approaching $t\sim 0.96 t_{\rm c}$, $\rho_\text{newt, central}$ decreases due to the bounce of the matter at the centre. As we have seen in Fig.\,\ref{fig:bounce8}, the density is not constant, and to approximate the GR solution better, we define a mean density:
\beq
\bar{\rho}_{\rm newt}=\frac{3}{4} \frac{M_\text{encl}}{ R_\text{equiv}^3}\,.
\eeq
For early times, this $ \bar{\rho}_{\rm newt}$ is identical to the central density, but the closer the simulations come to the bounce, the more it deviates. It reaches its maximum around $t\sim t_{\rm c}$, indicating that the bulk of the matter has bounced by that time. 

The approximated relativistic energy density has a different behaviour (solid black line in the figure). It shows two maxima just before and after $t=t_{\rm c}$. To understand this result, we show the two contributions to $\rho=\rho_1 + \rho_2$ explicitly. Here, $\rho_1=3 {\dot a}^2 \left({8\pi G a^2}\right)^{-1}$ is the kinetic contribution or the term related to the Hubble rate (dashed black line) and $\rho_2 = 3 k \left({8\pi G a^2}\right)^{-1}$ (dot\textendash dashed, black line) is related to the curvature. In contrast to the other curves, $\rho_1(t)$ has two peaks. This is caused by the Hubble rate, which follows the dynamics of the collapse. Initially, the cloud collapses slowly, so $H$ is small and negative. The collapse accelerates until the bounce of the very central matter sets in, causing $H$ to slowly decrease to a (negative) minimum. At this point, some of the matter changes direction and, therefore, the magnitude of the Hubble rate (and $\rho_1$) decreases. Around $t\sim t_{\rm c}$, $H$ and $\rho_1$ become zero, marking a critical point during the bounce. Subsequently, the matter continues to accelerate outwards and both $H$ and $\rho_1$ increase. However, there is still matter falling, causing the acceleration of the expansion to slow down until the (positive) maxima of $H$ and $\rho_1$ are reached. Finally, all matter has positive velocities and slows down during expansion. The amplitudes of $H$ and $\rho_1$ decrease. The magnitude of the negative minimum, corresponding to the infall velocity, is higher than that of the positive maximum, representing the bounce expansion. Consequently, the first peak of $\rho_1$ is larger than the second. 
At $t\sim t_{\rm c}$, when $\rho_1=0$, $\rho$ is dominated by $\rho_2$. This contribution behaves very similar to $\bar\rho_\text{newt}$ and has a maximum around $t\sim t_{\rm c}$.
That $\rho_2$ has its maximum right when $\rho_1=0$ in between its two maxima creates the two-peaked structure of $\rho$. 
Note that initially $\rho_2$ is larger than the Newtonian equivalents (pressureless solution or $\bar\rho_\text {newt}$), and it increases slower with time. This is caused by our choice of $k$ in the relativistic case. $k\neq0$ leads to the dependence $\rho_1\sim a^{-2}\sim R^{-2}$ in contrast to $\bar\rho_\text{newt} \sim R^{-3}$ for the Newtonian case. $\rho_1$ is also initially smaller than $\rho_2$. They intersect for the first time when $r = r_{\rm s}$. At this point $\rho_1=\rho_2$, which implies that $k = \dot{a}^2$ and consequently $\dot{r} = \dot{a}\chi = c$, which marks the formation of a BH in GR and shows that the bounce happens inside the BH. 

The different dependencies of $\bar\rho_\text{newt}$ and $\rho_2$ lead to very different asymptotic behaviour for very large $a$ (or $R$). This difference arises from the non-vanishing $k$ in the GR solution, which is a purely geometrical term absent in the Newtonian ($E=0$) solution. Only with this term, alongside the ground-state pressure (such as neutron degeneracy pressure), which causes the bounce in the Newtonian framework, can matter rebound in GR as well.
To see this, note how $P = -\rho$ in GR results in $\ddot{a} > 0$ (see equation\,\ref{eq:ddota}). This is a necessary but not sufficient condition for a bounce to occur. We also need $k > 0$ (in the equation\,\ref{eq:Friedmank}) to satisfy the additional bounce condition $\dot{a} = 0$ while $\rho > 0$.

In Fig.\,\ref{fig:bounce9}, we also plot the absolute value of the relativistic pressure. $P$ is equal to negative of the density ($P = -\rho$) on the two peaks of $\rho$. This happens because $\dot{\rho} = 0$ at these points and, consequently, $P(\rho) = -(\rho + \dot{\rho}/3H) = - \rho $.
Furthermore, when $H(t\sim t_{\rm c}) = 0$, the pressure diverges and $\rho_2$ and $\bar\rho_\text{newt}$ peak. This is what we expect from a speculated quantum degeneracy pressure: a bounce caused by a mechanism similar to the Pauli exclusion principle.

We may also interpret this behaviour with the help of the analogy to a relativistic scalar field $\psi$ in curved space\textendash time (see the Appendix \ref{sec:Quintessence}). 

When such a system is trapped in a ground state, which is defined by $V\gg K=\dot\psi^2$, the equations \eqref{eq:relat1} and \eqref{eq:relat2} read $P=K-V$ and $\rho=K+V$. Here, $K(\psi)$ and $V(\psi)$ are the kinetic and potential energies of the scalar field, respectively. When $V\gg K=\dot\psi^2$, the relativistic EoS can be expressed as $P=-\rho$. The pressure becoming equal to the negative energy density is exactly what we have found in Fig.\,\ref{fig:bounce9}. When $\rho$ has its two maxima, we find that $P=-\rho$. After the first peak of $\rho$, $P$ increases further to negative infinity when $H=0$. This huge negative pressure stops gravitational collapse and causes a rebound or bounce. Once the matter expands, the pressure reduces and tends towards $P=0$. During this phase, the curve passes again through $P=-\rho$ at the second peak in the relativistic density plot. 
The analogy with the scalar field cannot explain how to obtain $P< -\rho$ in between the two density maxima. However, we can see the similarity and consider the case when $P \leq -\rho$ as the ground state of our more complex system.

\subsection{BHU vs cosmological coupling}

This paper builds upon the theoretical foundations of the BHU model, and our results support its predictions. However, the BHU model is not the only alternative explanation for the accelerated expansion of the Universe. One compelling proposal is the concept of `cosmological coupling', which posits that certain objects, such as black holes (BHs), are coupled to the large-scale dynamics of the Universe \citep{McVittie:1933zz, Faraoni:2007es, Croker:2019mup, Croker:2021duf}.

Recent work by \cite{Farrah_2023} presents observational evidence that black holes can grow in mass without consuming gas or stars. This suggests that black holes might be a significant source of dark energy. Specifically, the study found that the mass of supermassive black holes at the centres of elliptical galaxies increases in a manner consistent with cosmological coupling. As the Universe expands, the mass of these black holes grows proportionately, contributing to the overall energy density of the Universe. Furthermore, \cite{Faraoni_2024} provides theoretical support for cosmological coupling by demonstrating that an exactly static BH event horizon cannot be embedded in a time-dependent geometry like the FLRW metric, inherently linking BHs to the Universe's expansion.

However, if black holes form through the spherical collapse process depicted in Fig.\,\ref{fig:collapse}, they create an empty transition region between the black hole (static only outside the Schwarzschild radius, $r_{\rm s}$) and the surrounding FLRW metric. According to the corollary of Birkhoff's theorem, this region would follow the Minkowski metric. Such a transition would effectively decouple the black hole from the FLRW metric, analogous to how regular perturbations, such as dark matter haloes, become virialized or collapse without contributing to dark energy coupling or global background dynamics. Additionally, \cite{Cadoni:2023lum} argues that non-singular black holes or horizonless compact objects in GR can exhibit cosmological coupling but are unlikely to act as a source of dark energy. This conclusion was supported by analysing the redshift dependence of mass growth in supermassive black holes within a sample of elliptical galaxies and comparing their theoretical predictions with higher redshift mass measurements obtained from the \textit{JWST}.

While both the BHU and cosmological coupling models challenge conventional theories and provide fresh perspectives on the role of black holes in the Universe's expansion, they offer distinct approaches and implications. The BHU model proposes that the Universe itself can be described as a black hole, obviating the need for dark energy by suggesting that the expansion of the Universe, driven by a non-singular bounce, behaves like a cosmological constant. This framework is consistent with an asymptotically de Sitter interior metric that smoothly transitions into an exterior Schwarzschild metric without discontinuities. In contrast, the cosmological coupling model attributes the mass growth of black holes to their interaction with the Universe's expansion, suggesting that this growth contributes to dark energy, which drives accelerated cosmic expansion.

Interestingly, our BHU could be one of many black holes within a larger manifold, potentially embedded in an expanding or contracting FLRW metric. This external metric could possess its own $\Lambda$ or dark energy term, different from the one measured inside our BHU, which is given by $\Lambda \simeq 3/r_{\rm s}^2$. If the cosmological coupling between our BHU and the external FLRW metric were a significant effect, it could influence our BHU mass and, consequently, the associated local term $\Lambda \simeq 3/r_{\rm s}^2$. In this scenario, our local $\Lambda$ would evolve over time, either decreasing or increasing (corresponding to an effective $\omega_0 < -1$), depending on whether the external FLRW metric expands or contracts. Such a framework provides a potential explanation for the recent evidence of $\omega_0 < -1$ reported by DESI2024 measurements (\citealt{2024arXiv240403002D}), without necessitating the introduction of ghost-like dark energy models.

\subsection{Potential signatures in the CMB}

Although the background equations for the big bang model and the BHU are identical\textemdash since the $\Lambda_{\rm EH}$ (or event horizon boundary) lies beyond the observable Universe\textemdash differences may arise in the origin and evolution of perturbations, as detailed in \cite{Camacho_2022}. Specifically, the size of super-horizon perturbations could be constrained by the BHU boundary during collapse. This constraint might explain observed anomalies in the CMB, such as the lower quadrupole and octopole moments, or the absence of perturbations on angular scales larger than 60 deg.

It is important to note that this effect is distinct from the parity asymmetry observed at $\ell < 20$, which hints at quantum effects on super-horizon scales and could also provide an explanation for the anomalies in the low quadrupole and hemispherical power asymmetry (\citealt{2024JCAP...06..001G}).

In the standard model of inflation, super-horizon perturbations are explained as being "frozen" when they cross the horizon during the inflationary phase, only to re-enter at a later stage of the Universe’s evolution. Although the BHU model solves the horizon problem through a similar causal mechanism, it introduces a key difference: a cutoff exists for scales larger than $\lambda > 2R \ (k < \pi/R)$ \citep{Gaztanaga:bhu}, unlike the scale-invariant spectrum predicted by inflation. This distinction results in the anomalous absence of the largest structures in the CMB sky temperature (on top of the quantum parity asymmetry mentioned before), which potentially explains the observed deviations at these scales.

Furthermore, perturbations generated during the bounce phase of the BHU model could lead to notable consequences. For example (see \citealt{Gaztanaga:bhu}):

- A portion or even the entirety of dark matter might consist of primordial black holes or primordial neutron stars formed during the bounce.

- The bounce mechanism could offer insights into the measured entropy of the Universe, particularly the observed baryon-to-photon ratio.

These implications underscore the need for further investigation, which forms the central motivation for the research presented in this paper.

\section{Conclusion}\label{sec:conclusions}

This work is motivated by finding a scenario which has the potential to explain various features of the behaviour of our Universe (e.g. the singularity, flatness, dark energy, and horizon problems), which cannot be solved easily within the standard model of cosmology. Following the BHU model \citep{Gaztanaga:bhu1,Gaztanaga:bhu2}, our Universe is inside a black hole and is currently in a state of expansion. Before this expansion, the Universe is thought to have collapsed from a cold, tenuous cloud of matter. Once a critical density had been reached, this matter had to bounce back. As a first step to understanding the processes and necessary conditions of such a bouncing universe model, we investigated the collapse of a spherically symmetric matter distribution and how a potential bounce can occur for a simplistic setup.

We started by discussing the general relativistic equations that describe the collapse of a uniform spherical cloud. For a cloud with zero pressure and zero total energy, these equations are mathematically equivalent to a Newtonian description. This equivalence is a well-known fact in GR. A numerical simulation of one such cloud in Newtonian hydrodynamics with a mass of $8\,\rm M_\odot$ and $P=0$ shows that the radius and density evolve approximately as predicted by the analytical GR solution. Moreover, this cloud remains approximately FLRW (uniform density) at all times. 
In our approach, we were able to simplify the model such that the formation of a BH, a genuine GR effect, can be inferred from a Newtonian simulation, significantly easing the numerical implementation. This simplification is valid only under the assumption of spherical symmetry and when vorticity and tidal fields are neglected \citep[see \S\ref{sec:SC} and][]{faraoni}. Proceeding with the Newtonian code and abandoning the approximation of zero pressure, we simulate the collapse of FLRW clouds of masses $5$, $20$ and $1000$ M$_\odot$ with a polytropic EoS. This is a simple EoS that creates a bounce and is motivated by its frequent use to describe neutron stars that result from a collapse of stars with masses from $8$ to $40$ M$_\odot$. Since this is a qualitative study, we do not intend to solve this collapse in all detail, but rather want to study the basic features on the route to understand the collapse of much more massive configurations. Our scenario of a cold collapse is significantly simpler than a stellar collapse, where all kinds of nuclear reactions and neutrino-matter interaction play important roles, and a much simpler barotropic EoS is justified here. However,
the exact equivalent GR solution for an EoS with non-zero pressure is not straightforward and is left to be done in future work.

For the cases with non-zero pressures, initially, the central densities increase in our simulations as in the pressureless solution. However, due to the equally increasing pressure, the matter bounces once it reaches a certain maximum density, and subsequently, the central density starts to decrease. Apart from the lightest model with the stiffest EoS, this happens when $95$ per cent of the total mass is already inside its Schwarzschild radius. This means that in the GR framework, a BH has formed, and matter collapses and bounces inside its event horizon. 
For the masses we simulated here, this repulsion leading to the bounce could be caused by neutron or even quark degeneracy. However, for much larger masses comparable to the mass of our Universe, which we finally aim at collapsing, none of the known physical effects could stop the collapse. In the BHU model, we thus assume that at some point, some supranuclear saturation densities will be reached, and an exotic form of quantum matter will instigate the bounce due to a quantum mechanism similar to Pauli's exclusion principle.
Crucially, this quantum mechanism violates the strong energy condition in classical GR and, together with $k>0$, sidesteps the singularity GR theorems proposed by Hawking and Penrose \cite{Hawking_Penrose_Singularity}, thus presenting a novel solution to a pivotal issue in cosmological theory. 

Fig.\,\ref{fig:bounce9} illustrates our key findings. 
When estimating the properties of the relativistic bounce with our Newtonian simulations with a polytropic EoS, we find conditions where $P=-\rho$. This usually occurs when a ground state of the matter is reached (see the Appendix \ref{sec:Quintessence} for an example of a scalar field). As stated before, such a ground state could be characterized by the Pauli Exclusion Principle for an unknown form of quantum matter. 

We started our investigation from the equivalence of the pressureless collapse in GR and Newtonian descriptions and used a Newtonian simulation with pressure to estimate what one can expect in the full GR case. We have focused on distributions containing the order of magnitude of stellar size masses up to $1000$ M$_\odot$, where the pressure can come from the neutron degeneracy pressure. For larger masses, the particular mechanism that creates the required extreme pressures is highly speculative. However, from quantum principles, it seems reasonable that matter cannot be compressed arbitrarily. As a next step, we intend to increase the simulated masses and generalize our results by treating the entire collapse within GR. This includes, in particular, the requirement to drop the condition of homogeneous density and pressure in the GR case.

\newpage
\section*{Acknowledgments}
We thank the anonymous referee for their valuable comments.
Enrique Gazta\~naga acknowledges Sravan Kumar for pointing out the important role of curvature  and grants from Spain Plan Nacional (PGC2018-102021-B-100) and
Maria de Maeztu (CEX2020-001058-M). Swaraj Pradhan acknowledges the support and hospitality at the Institute of Cosmology and Gravitation during a 3-month stay to conclude his Master's Thesis.
Michael Gabler acknowledges the support through the Generalitat Valenciana via the grant CIDEGENT/2019/031, 
the grant PID2021-127495NB-I00 funded by MCIN/AEI/10.13039/501100011033 and by the European Union, and the Astrophysics and High Energy Physics programme of the Generalitat Valenciana ASFAE/2022/026 funded by MCIN and the European Union NextGenerationEU (PRTR-C17.I1). 

\section*{Data Availability}

The main modification to the \textsc{CASTRO} code in the form of a cold collapse test problem, together with a Python notebook to perform the data analysis, can be found in the following GitHub repository compiled by the authors:
\url{https://github.com/JAR-AWS/BHU_MODEL}

\bibliographystyle{mnras}
\bibliography{bibliography}

\begin{thebibliography}{}
\makeatletter
\relax
\def\mn@urlcharsother{\let\do\@makeother \do\$\do\&\do\#\do\^\do\_\do\%\do\~}
\def\mn@doi{\begingroup\mn@urlcharsother \@ifnextchar [ {\mn@doi@} {\mn@doi@[]}}
\def\mn@doi@[#1]#2{\def\@tempa{#1}\ifx\@tempa\@empty \href {http://dx.doi.org/#2} {doi:#2}\else \href {http://dx.doi.org/#2} {#1}\fi \endgroup}
\def\mn@eprint#1#2{\mn@eprint@#1:#2::\@nil}
\def\mn@eprint@arXiv#1{\href {http://arxiv.org/abs/#1} {{\tt arXiv:#1}}}
\def\mn@eprint@dblp#1{\href {http://dblp.uni-trier.de/rec/bibtex/#1.xml} {dblp:#1}}
\def\mn@eprint@#1:#2:#3:#4\@nil{\def\@tempa {#1}\def\@tempb {#2}\def\@tempc {#3}\ifx \@tempc \@empty \let \@tempc \@tempb \let \@tempb \@tempa \fi \ifx \@tempb \@empty \def\@tempb {arXiv}\fi \@ifundefined {mn@eprint@\@tempb}{\@tempb:\@tempc}{\expandafter \expandafter \csname mn@eprint@\@tempb\endcsname \expandafter{\@tempc}}}

\bibitem[\protect\citeauthoryear{{Aguirre} \& {Johnson}}{{Aguirre} \& {Johnson}}{2005}]{Aguirre}
{Aguirre} A.,  {Johnson} M.~C.,  2005, PRD, 72, 103525

\bibitem[\protect\citeauthoryear{Alcubierre, Brügmann, Pollney, Seidel  \& Takahashi}{Alcubierre et~al.}{2001}]{Alcubierre_2001}
Alcubierre M.,  Brügmann B.,  Pollney D.,  Seidel E.,   Takahashi R.,  2001, \mn@doi [Physical Review D] {10.1103/physrevd.64.061501}, 64

\bibitem[\protect\citeauthoryear{Almgren et~al.,}{Almgren et~al.}{2010}]{Castro1}
Almgren A.~S.,  et~al., 2010, \mn@doi [The Astrophysical Journal] {10.1088/0004-637x/715/2/1221}, 715, 1221–1238

\bibitem[\protect\citeauthoryear{Baker, Centrella, Choi, Koppitz  \& van Meter}{Baker et~al.}{2006}]{Baker_2006_puncture}
Baker J.~G.,  Centrella J.,  Choi D.-I.,  Koppitz M.,   van Meter J.,  2006, \mn@doi [Physical Review Letters] {10.1103/physrevlett.96.111102}, 96, 111102

\bibitem[\protect\citeauthoryear{{Barrow}}{{Barrow}}{1988}]{Barrow_SEC}
{Barrow} J.~D.,  1988, \qjras, \href {https://ui.adsabs.harvard.edu/abs/1988QJRAS..29..101B} {29, 101}

\bibitem[\protect\citeauthoryear{{Bernardeau}, {Colombi}, {Gazta{\~n}aga}  \& {Scoccimarro}}{{Bernardeau} et~al.}{2002}]{Bernardeau}
{Bernardeau} F.,  {Colombi} S.,  {Gazta{\~n}aga} E.,   {Scoccimarro} R.,  2002, PhysRep, \href {https://ui.adsabs.harvard.edu/abs/2002PhR...367....1B} {367, 1}

\bibitem[\protect\citeauthoryear{{Bethe}}{{Bethe}}{1990}]{bethe1990}
{Bethe} H.~A.,  1990, \mn@doi [Reviews of Modern Physics] {10.1103/RevModPhys.62.801}, \href {https://ui.adsabs.harvard.edu/abs/1990RvMP...62..801B} {62, 801}

\bibitem[\protect\citeauthoryear{{Betoule, M.} et~al.,}{{Betoule, M.} et~al.}{2014}]{Betoule_2012}
{Betoule, M.} et~al., 2014, \mn@doi [A&A] {10.1051/0004-6361/201423413}, 568, A22

\bibitem[\protect\citeauthoryear{{Blau}, {Guendelman}  \& {Guth}}{{Blau} et~al.}{1987}]{1987PhRvD..35.1747B}
{Blau} S.~K.,  {Guendelman} E.~I.,   {Guth} A.~H.,  1987, PRD, 35, 1747

\bibitem[\protect\citeauthoryear{{Bondi}}{{Bondi}}{1969}]{bondi}
{Bondi} H.,  1969, MNRAS, 142, 333

\bibitem[\protect\citeauthoryear{Brandenberger \& Peter}{Brandenberger \& Peter}{2017}]{Brandenberger_bouncing}
Brandenberger R.,  Peter P.,  2017, \mn@doi [Foundations of Physics] {10.1007/s10701-016-0057-0}, 47, 797–850

\bibitem[\protect\citeauthoryear{{Br{\"u}gmann}, {Gonz{\'a}lez}, {Hannam}, {Husa}, {Sperhake}  \& {Tichy}}{{Br{\"u}gmann} et~al.}{2008}]{Brugmann_puncture}
{Br{\"u}gmann} B.,  {Gonz{\'a}lez} J.~A.,  {Hannam} M.,  {Husa} S.,  {Sperhake} U.,   {Tichy} W.,  2008, \mn@doi [\prd] {10.1103/PhysRevD.77.024027}, \href {https://ui.adsabs.harvard.edu/abs/2008PhRvD..77b4027B} {77, 024027}

\bibitem[\protect\citeauthoryear{Burd \& Barrow}{Burd \& Barrow}{1988}]{Burd:1988ss_SEC}
Burd A.~B.,  Barrow J.~D.,  1988, \mn@doi [Nucl. Phys. B] {10.1016/0550-3213(88)90135-6}, 308, 929

\bibitem[\protect\citeauthoryear{{Burrows}, {Hayes}  \& {Fryxell}}{{Burrows} et~al.}{1995}]{burrows1995}
{Burrows} A.,  {Hayes} J.,   {Fryxell} B.~A.,  1995, \mn@doi [\apj] {10.1086/176188}, 450, 830

\bibitem[\protect\citeauthoryear{Cadoni, Sanna, Pitzalis, Banerjee, Murgia, Hazra  \& Branchesi}{Cadoni et~al.}{2023}]{Cadoni:2023lum}
Cadoni M.,  Sanna A.~P.,  Pitzalis M.,  Banerjee B.,  Murgia R.,  Hazra N.,   Branchesi M.,  2023, \mn@doi [JCAP] {10.1088/1475-7516/2023/11/007}, 11, 007

\bibitem[\protect\citeauthoryear{Campanelli, Lousto, Marronetti  \& Zlochower}{Campanelli et~al.}{2006}]{Campanelli_2006_puncture}
Campanelli M.,  Lousto C.~O.,  Marronetti P.,   Zlochower Y.,  2006, \mn@doi [Physical Review Letters] {10.1103/physrevlett.96.111101}, 96, 111101

\bibitem[\protect\citeauthoryear{Chandrasekhar}{Chandrasekhar}{1931}]{Chandrasekhar}
Chandrasekhar S.,  1931, \mn@doi [Astrophys. J.] {10.1086/143324}, 74, 81

\bibitem[\protect\citeauthoryear{Croker \& Weiner}{Croker \& Weiner}{2019}]{Croker:2019mup}
Croker K.~S.,  Weiner J.~L.,  2019, \mn@doi [Astrophys. J.] {10.3847/1538-4357/ab32da}, 882, 19

\bibitem[\protect\citeauthoryear{Croker, Zevin, Farrah, Nishimura  \& Tarle}{Croker et~al.}{2021}]{Croker:2021duf}
Croker K.~S.,  Zevin M.~J.,  Farrah D.,  Nishimura K.~A.,   Tarle G.,  2021, \mn@doi [Astrophys. J. Lett.] {10.3847/2041-8213/ac2fad}, 921, L22

\bibitem[\protect\citeauthoryear{{DESI Collaboration} et~al.,}{{DESI Collaboration} et~al.}{2024}]{2024arXiv240403002D}
{DESI Collaboration} et~al., 2024, \mn@doi [arXiv e-prints] {10.48550/arXiv.2404.03002}, \href {https://ui.adsabs.harvard.edu/abs/2024arXiv240403002D} {p. arXiv:2404.03002}

\bibitem[\protect\citeauthoryear{Daghigh, Kapusta  \& Hosotani}{Daghigh et~al.}{2000}]{BH_interior_daghigh}
Daghigh R.~G.,  Kapusta J.~I.,   Hosotani Y.,  2000, False Vacuum Black Holes and Universes (\mn@eprint {arXiv} {gr-qc/0008006})

\bibitem[\protect\citeauthoryear{{Dodelson}}{{Dodelson}}{2003}]{Dodelson}
{Dodelson} S.,  2003, {Modern cosmology, Academic Press, NY}

\bibitem[\protect\citeauthoryear{Duez, Shapiro  \& Yo}{Duez et~al.}{2004}]{Duez2004_excision}
Duez M.~D.,  Shapiro S.~L.,   Yo H.-J.,  2004, \mn@doi [Phys. Rev. D] {10.1103/PhysRevD.69.104016}, 69, 104016

\bibitem[\protect\citeauthoryear{Eisenstein et~al.,}{Eisenstein et~al.}{2005}]{Eisenstein_2005}
Eisenstein D.~J.,  et~al., 2005, \mn@doi [The Astrophysical Journal] {10.1086/466512}, 633, 560

\bibitem[\protect\citeauthoryear{Escriv\`a}{Escriv\`a}{2020}]{Escriva:spribhos}
Escriv\`a A.,  2020, \mn@doi [Phys. Dark Univ.] {10.1016/j.dark.2020.100466}, 27, 100466

\bibitem[\protect\citeauthoryear{Faraoni \& Atieh}{Faraoni \& Atieh}{2020}]{faraoni}
Faraoni V.,  Atieh F.,  2020, \mn@doi [Phys. Rev. D] {10.1103/PhysRevD.102.044020}, 102, 044020

\bibitem[\protect\citeauthoryear{Faraoni \& Jacques}{Faraoni \& Jacques}{2007}]{Faraoni:2007es}
Faraoni V.,  Jacques A.,  2007, \mn@doi [Phys. Rev. D] {10.1103/PhysRevD.76.063510}, 76, 063510

\bibitem[\protect\citeauthoryear{Faraoni \& Rinaldi}{Faraoni \& Rinaldi}{2024}]{Faraoni_2024}
Faraoni V.,  Rinaldi M.,  2024, \mn@doi [Phys. Rev. D] {10.1103/PhysRevD.110.063553}, 110, 063553

\bibitem[\protect\citeauthoryear{{Farhi} \& {Guth}}{{Farhi} \& {Guth}}{1987}]{Farhi}
{Farhi} E.,  {Guth} A.~H.,  1987, Physics Letters B, 183, 149

\bibitem[\protect\citeauthoryear{Farrah et~al.,}{Farrah et~al.}{2023}]{Farrah_2023}
Farrah D.,  et~al., 2023, \mn@doi [The Astrophysical Journal Letters] {10.3847/2041-8213/acb704}, 944, L31

\bibitem[\protect\citeauthoryear{{Ferreira} \& {Provid{\^e}ncia}}{{Ferreira} \& {Provid{\^e}ncia}}{2021}]{Ferreira2021}
{Ferreira} M.,  {Provid{\^e}ncia} C.,  2021, \mn@doi [\prd] {10.1103/PhysRevD.104.063006}, \href {https://ui.adsabs.harvard.edu/abs/2021PhRvD.104f3006F} {104, 063006}

\bibitem[\protect\citeauthoryear{{Foglizzo}}{{Foglizzo}}{2017}]{foglizzo2017}
{Foglizzo} T.,  2017, in {Alsabti} A.~W.,  {Murdin} P.,  eds, , Handbook of Supernovae.
Springer International Publishing AG, p.~1053, \mn@doi{10.1007/978-3-319-21846-5_52}

\bibitem[\protect\citeauthoryear{{Fosalba}, {Gazta{\~n}aga}  \& {Castander}}{{Fosalba} et~al.}{2003}]{2003ApJ...597L..89F}
{Fosalba} P.,  {Gazta{\~n}aga} E.,   {Castander} F.~J.,  2003, \mn@doi [ApJL] {10.1086/379848}, \href {https://ui.adsabs.harvard.edu/abs/2003ApJ...597L..89F} {597, L89}

\bibitem[\protect\citeauthoryear{{Frolov}, {Markov}  \& {Mukhanov}}{{Frolov} et~al.}{1989}]{1989PhLB..216..272F}
{Frolov} V.~P.,  {Markov} M.~A.,   {Mukhanov} V.~F.,  1989, Phys Let B, 216, 272

\bibitem[\protect\citeauthoryear{{Gazta{\~n}aga}}{{Gazta{\~n}aga}}{2021}]{Gaztanaga:2021lamb1}
{Gazta{\~n}aga} E.,  2021, \mn@doi [MNRAS] {10.1093/mnras/stab056}, 502, 436

\bibitem[\protect\citeauthoryear{{Gazta{\~n}aga}}{{Gazta{\~n}aga}}{2022a}]{Gaztanaga:bhu}
{Gazta{\~n}aga} E.,  2022a, \mn@doi [Universe] {10.3390/universe8050257}, 8, 257

\bibitem[\protect\citeauthoryear{{Gazta{\~n}aga}}{{Gazta{\~n}aga}}{2022b}]{Gaztanaga:2022lamb2}
{Gazta{\~n}aga} E.,  2022b, \mn@doi [Symmetry] {10.3390/sym14020300}, 14, 300

\bibitem[\protect\citeauthoryear{{Gazta{\~n}aga}}{{Gazta{\~n}aga}}{2022c}]{Gaztanaga:bhu1}
{Gazta{\~n}aga} E.,  2022c, \mn@doi [Symmetry] {10.3390/sym14091849}, 14, 1849

\bibitem[\protect\citeauthoryear{{Gazta{\~n}aga}}{{Gazta{\~n}aga}}{2022d}]{Gaztanaga:bhu2}
{Gazta{\~n}aga} E.,  2022d, \mn@doi [Symmetry] {10.3390/sym14101984}, 14, 1984

\bibitem[\protect\citeauthoryear{{Gazta{\~n}aga}}{{Gazta{\~n}aga}}{2023a}]{whiteholes}
{Gazta{\~n}aga} E.,  2023a, \mn@doi [Universe] {10.3390/universe9040194}, \href {https://ui.adsabs.harvard.edu/abs/2023Univ....9..194G} {9, 194}

\bibitem[\protect\citeauthoryear{{Gazta{\~n}aga}}{{Gazta{\~n}aga}}{2023b}]{gaztanaga_mou}
{Gazta{\~n}aga} E.,  2023b, \mn@doi [MNRAS] {10.1093/mnrasl/slad015}, \href {https://ui.adsabs.harvard.edu/abs/2023MNRAS.521L..59G} {521, L59}

\bibitem[\protect\citeauthoryear{{Gazta{\~n}aga}}{{Gazta{\~n}aga}}{2024}]{decceleration}
{Gazta{\~n}aga} E.,  2024, \mn@doi [Symmetry] {10.3390/sym16091141}, \href {https://ui.adsabs.harvard.edu/abs/2024Symm...16.1141G} {16, 1141}

\bibitem[\protect\citeauthoryear{{Gazta{\~n}aga} \& {Camacho-Quevedo}}{{Gazta{\~n}aga} \& {Camacho-Quevedo}}{2022}]{Camacho_2022}
{Gazta{\~n}aga} E.,  {Camacho-Quevedo} B.,  2022, \mn@doi [Physics Letters B] {10.1016/j.physletb.2022.137468}, \href {https://ui.adsabs.harvard.edu/abs/2022PhLB..83537468G} {835, 137468}

\bibitem[\protect\citeauthoryear{{Gazta{\~n}aga} \& {Sravan Kumar}}{{Gazta{\~n}aga} \& {Sravan Kumar}}{2024}]{2024JCAP...06..001G}
{Gazta{\~n}aga} E.,  {Sravan Kumar} K.,  2024, \mn@doi [\jcap] {10.1088/1475-7516/2024/06/001}, \href {https://ui.adsabs.harvard.edu/abs/2024JCAP...06..001G} {2024, 001}

\bibitem[\protect\citeauthoryear{{Gibbons} \& {Hawking}}{{Gibbons} \& {Hawking}}{1977}]{Gibbons}
{Gibbons} G.~W.,  {Hawking} S.~W.,  1977, PRD, 15, 2738

\bibitem[\protect\citeauthoryear{{Gonzalez-Diaz}}{{Gonzalez-Diaz}}{1981}]{BH_interior_Gonzalez}
{Gonzalez-Diaz} P.~F.,  1981, Nuovo Cimento Lettere, \href {https://ui.adsabs.harvard.edu/abs/1981NCimL......161G} {32, 161}

\bibitem[\protect\citeauthoryear{{Gr{\o}n} \& {Soleng}}{{Gr{\o}n} \& {Soleng}}{1989}]{1989PhLA..138...89G}
{Gr{\o}n} {\O}.,  {Soleng} H.~H.,  1989, Physics Letters A, 138, 89

\bibitem[\protect\citeauthoryear{{Hawking} \& {Horowitz}}{{Hawking} \& {Horowitz}}{1996}]{Hawking1996}
{Hawking} S.~W.,  {Horowitz} G.~T.,  1996, Class Quantum Gravity, 13, 1487

\bibitem[\protect\citeauthoryear{Hawking, Penrose  \& Bondi}{Hawking et~al.}{1970}]{Hawking_Penrose_Singularity}
Hawking S.~W.,  Penrose R.,   Bondi H.,  1970, \mn@doi [Proceedings of the Royal Society of London. A. Mathematical and Physical Sciences] {10.1098/rspa.1970.0021}, 314, 529

\bibitem[\protect\citeauthoryear{Hebeler, Lattimer, Pethick  \& Schwenk}{Hebeler et~al.}{2013}]{Hebeler_2013}
Hebeler K.,  Lattimer J.~M.,  Pethick C.~J.,   Schwenk A.,  2013, \mn@doi [The Astrophysical Journal] {10.1088/0004-637X/773/1/11}, 773, 11

\bibitem[\protect\citeauthoryear{{Janka}}{{Janka}}{2017}]{janka2017}
{Janka} H.-T.,  2017, in {Alsabti} A.~W.,  {Murdin} P.,  eds, , Handbook of Supernovae.
Springer International Publishing AG, p.~1095, \mn@doi{10.1007/978-3-319-21846-5_109}

\bibitem[\protect\citeauthoryear{{Janka}, {Hanke}, {H{\"u}depohl}, {Marek}, {M{\"u}ller}  \& {Obergaulinger}}{{Janka} et~al.}{2012}]{janka2012}
{Janka} H.-T.,  {Hanke} F.,  {H{\"u}depohl} L.,  {Marek} A.,  {M{\"u}ller} B.,   {Obergaulinger} M.,  2012, \mn@doi [Progress of Theoretical and Experimental Physics] {10.1093/ptep/pts067}, \href {https://ui.adsabs.harvard.edu/abs/2012PTEP.2012aA309J} {2012, 01A309}

\bibitem[\protect\citeauthoryear{Jimenez \& Loeb}{Jimenez \& Loeb}{2002}]{Jimenez_2002}
Jimenez R.,  Loeb A.,  2002, \mn@doi [The Astrophysical Journal] {10.1086/340549}, 573, 37

\bibitem[\protect\citeauthoryear{Johansen \& Ravndal}{Johansen \& Ravndal}{2006}]{BirkhoffH}
Johansen N.,  Ravndal F.,  2006, General Relativity and Gravitation, 38, 537

\bibitem[\protect\citeauthoryear{Kothari \& Saha}{Kothari \& Saha}{1938}]{kothari_pressure_ionization}
Kothari D.~S.,  Saha M.~N.,  1938, \mn@doi [Proceedings of the Royal Society of London. Series A. Mathematical and Physical Sciences] {10.1098/rspa.1938.0073}, 165, 486

\bibitem[\protect\citeauthoryear{{Kusenko}}{{Kusenko}}{2020}]{PBH3}
{Kusenko} A.~e.,  2020, PRL, 125, 181304

\bibitem[\protect\citeauthoryear{{Landau} \& {Lifshitz}}{{Landau} \& {Lifshitz}}{1971}]{Landau1971}
{Landau} L.~D.,  {Lifshitz} E.~M.,  1971, {The classical theory of fields}.
Pergamon Press, New York

\bibitem[\protect\citeauthoryear{{Lattimer} \& {Prakash}}{{Lattimer} \& {Prakash}}{2007}]{Lattimer_Prakash_eos}
{Lattimer} J.~M.,  {Prakash} M.,  2007, \mn@doi [PhysRep] {10.1016/j.physrep.2007.02.003}, \href {https://ui.adsabs.harvard.edu/abs/2007PhR...442..109L} {442, 109}

\bibitem[\protect\citeauthoryear{{Lema{\^\i}tre}}{{Lema{\^\i}tre}}{1927}]{Lemaitre}
{Lema{\^\i}tre} G.,  1927, Annales de la S.S. de Bruxelles, \href {https://ui.adsabs.harvard.edu/abs/1927ASSB...47...49L} {47, 49}

\bibitem[\protect\citeauthoryear{{Limongi}}{{Limongi}}{2017}]{limongi2017}
{Limongi} M.,  2017, in {Alsabti} A.~W.,  {Murdin} P.,  eds, , Handbook of Supernovae.
Springer International Publishing AG, p.~513, \mn@doi{10.1007/978-3-319-21846-5_119}

\bibitem[\protect\citeauthoryear{Malafarina}{Malafarina}{2018}]{bounce_rev1}
Malafarina D.,  2018, \mn@doi [Universe] {10.3390/universe4090092}, 4

\bibitem[\protect\citeauthoryear{Marsa \& Choptuik}{Marsa \& Choptuik}{1996}]{Marsa_Choptuik_excision}
Marsa R.~L.,  Choptuik M.~W.,  1996, \mn@doi [Phys. Rev. D] {10.1103/PhysRevD.54.4929}, 54, 4929

\bibitem[\protect\citeauthoryear{Mazur \& Mottola}{Mazur \& Mottola}{2015}]{gravastar2015}
Mazur P.~O.,  Mottola E.,  2015, Classical and Quantum Gravity, 32, 215024

\bibitem[\protect\citeauthoryear{McVittie}{McVittie}{1933}]{McVittie:1933zz}
McVittie G.~C.,  1933, \mn@doi [Mon. Not. Roy. Astron. Soc.] {10.1093/mnras/93.5.325}, 93, 325

\bibitem[\protect\citeauthoryear{Misner \& Sharp}{Misner \& Sharp}{1964}]{Misner-Sharp}
Misner C.~W.,  Sharp D.~H.,  1964, \mn@doi [Phys. Rev.] {10.1103/PhysRev.136.B571}, 136, B571

\bibitem[\protect\citeauthoryear{{Monchmeyer} \& {Muller}}{{Monchmeyer} \& {Muller}}{1989}]{mnm}
{Monchmeyer} R.,  {Muller} E.,  1989, Astronomy and Astrophysics, 217, 351

\bibitem[\protect\citeauthoryear{Montero, Janka  \& Müller}{Montero et~al.}{2012}]{Montero_2012}
Montero P.~J.,  Janka H.-T.,   Müller E.,  2012, \mn@doi [The Astrophysical Journal] {10.1088/0004-637x/749/1/37}, 749, 37

\bibitem[\protect\citeauthoryear{Olmedo, Saini  \& Singh}{Olmedo et~al.}{2017}]{Olmedo_QG}
Olmedo J.,  Saini S.,   Singh P.,  2017, \mn@doi [Classical and Quantum Gravity] {10.1088/1361-6382/aa8da8}, 34, 225011

\bibitem[\protect\citeauthoryear{Oppenheimer \& Snyder}{Oppenheimer \& Snyder}{1939}]{Oppenheimer-Snyder}
Oppenheimer J.~R.,  Snyder H.,  1939, \mn@doi [Phys. Rev.] {10.1103/PhysRev.56.455}, 56, 455

\bibitem[\protect\citeauthoryear{{Padmanabhan}}{{Padmanabhan}}{2010}]{padmanabhan_2010}
{Padmanabhan} T.,  2010, {Gravitation, Cambridge Univ. Press}

\bibitem[\protect\citeauthoryear{Perlmutter et~al.,}{Perlmutter et~al.}{1999}]{Perlmutter_1999}
Perlmutter S.,  et~al., 1999, \mn@doi [The Astrophysical Journal] {10.1086/307221}, 517, 565

\bibitem[\protect\citeauthoryear{{Riess} et~al.,}{{Riess} et~al.}{1998}]{Riess_1998}
{Riess} A.~G.,  et~al., 1998, \mn@doi [\aj] {10.1086/300499}, \href {https://ui.adsabs.harvard.edu/abs/1998AJ....116.1009R} {116, 1009}

\bibitem[\protect\citeauthoryear{Rovelli}{Rovelli}{2007}]{Rovell_QG}
Rovelli C.,  2007, Quantum Gravity.
Cambridge University Press

\bibitem[\protect\citeauthoryear{Schmidt et~al.,}{Schmidt et~al.}{1998}]{Schmidt_1998}
Schmidt B.~P.,  et~al., 1998, \mn@doi [The Astrophysical Journal] {10.1086/306308}, 507, 46

\bibitem[\protect\citeauthoryear{Seidel \& Suen}{Seidel \& Suen}{1992}]{Seidel_excision}
Seidel E.,  Suen W.-M.,  1992, \mn@doi [Phys. Rev. Lett.] {10.1103/PhysRevLett.69.1845}, 69, 1845

\bibitem[\protect\citeauthoryear{{Tolman}}{{Tolman}}{1934}]{Tolman1934}
{Tolman} R.~C.,  1934, Proc. of the Nat. Academy of Science, \href {https://ui.adsabs.harvard.edu/abs/1934PNAS...20..169T} {20, 169}

\bibitem[\protect\citeauthoryear{{Weinberg}}{{Weinberg}}{2008}]{Cosmology_book_Weinberg}
{Weinberg} S.,  2008, {Cosmology, Oxford University Press}

\bibitem[\protect\citeauthoryear{{York}}{{York}}{1972}]{York}
{York} J.~W.,  1972, Phy.Rev.Lett., 28, 1082

\makeatother
\end{thebibliography}

\appendix 

\section{Relativist and Newtonian Mass}
\label{app:mass}

The relativistic Misner\textendash Sharp mass energy $M_{\rm MS}$ (\citealt{Misner-Sharp}) inside a spatial hypersurface $\Sigma$ of the space\textendash time determined by the metric, equation\,\eqref{eq:SphericalMetric2} for $r<R$ or $\chi<\chi_*$ is:

\beq
M_{\rm MS} = \int_0^R \rho 4\pi r^2 dr
= \int_0^{\chi_*} \rho \,\left(1 + \frac{\dot{r}^2}{c^2} - \frac{{2GM_{\rm MS}}}{c^2r} \right)^{1/2}  dV_3  
\label{eq:E}
\eeq
where $dV_3 = d^3y \sqrt{-h} = 4\pi r^2 e^{\lambda/2} \text{d}\chi$ is the 3D spatial volume element of the metric in $\Sigma$. Note that we recover here units if $c \ne 1$ to compare to the non-relativistic limit. The first term is the matter or Newtonian mass (which we call here $M$):
\beq
M =  \int_\Sigma\rho \,  dV_3   = \int_0^{\chi_*}  \rho \,  4\pi r^2 (\upartial_\chi r) \text{d}\chi
\label{eq:N_m}
\eeq
and corresponds to equation\,\eqref{eq:Mass}.
We can then interpret the next two terms in equation\,\eqref{eq:E} as the contribution to mass\textendash energy from the kinetic and potential energies, respectively. In the non-relativistic limit ($c=\infty$), these two terms are negligible, and $M_{\rm MS}=M$. However, in general, as indicated by equation\,\eqref{eq:E}, $M_{\rm MS}$ can not be expressed as a sum of individual energies as $M_{\rm MS}$ also appears inside the integral, reflecting the non-linear nature of gravity. But in the case of equation\,\eqref{eq:H2}, the kinetic and potential energy terms cancel for $M=M_{\rm MS}$, and we can interpret $M$ as the total relativistic mass-energy of the system.

\section{Scalar fields in curved space-time}
\label{sec:Quintessence}

To understand nuclear saturation density, it is essential to grasp how the Pauli exclusion principle leads to a ground state of matter. As mentioned in the introduction, we do not yet have a full understanding of this process in terms of fundamental physics for massive stars or even more massive mass distributions. 
To build intuition about the relativistic EoS and the behaviour of relativistic matter\textendash energy as it reaches the ground state, a useful approach is to study the analogies to a system having a scalar field \(\psi(r)\) as an effective degree of freedom. The key here is to account for relativistic effects by placing the field in a curved space\textendash time. By deriving the energy\textendash momentum tensor \(\boldsymbol{T}_{\mu\nu}\) for this scalar field system and comparing it to that of a relativistic perfect fluid, we gain better insight into the equivalent EoS for the ground state of $\psi$.

Consider the Einstein\textendash Hilbert action with minimally coupled matter fields with Lagrangian ${\cal L}$ (Eq.93.2 and 95.4 in \cite{Landau1971}):
\beq
S = \int_{\calMa} \text{d}\calMa \left[ \frac{ R- 2 \Lambda}{16\pi G} +  {\cal L} \right]\,,
\label{eq:actionC}
\eeq
where the energy\textendash momentum $\boldsymbol{T}_{\mu\nu}$ is defined as:
\beq
\boldsymbol{T}_{\mu\nu} \equiv -{2\over{\sqrt{g}}} {\delta (\sqrt{-g}  {\cal L}) \over{\delta \boldsymbol{g}^{\mu\nu} }}
= \boldsymbol{g}_{\mu\nu} {\cal L}  - 2 {\upartial {\cal L} \over{\upartial \boldsymbol{g}^{\mu\nu}}} .
\label{eq:TmunuL}
\eeq
The least action principle with respect to the metric $\boldsymbol{g}_{\mu\nu}$
yields  Einstein Field equations:
\beq
{\delta S \over{\delta \boldsymbol{g}^{\mu\nu} }}=0 ~\rightarrow  ~
\boldsymbol{R}_{\mu\nu} - {1\over{2}} \boldsymbol{g}_{\mu\nu} R    + \Lambda \boldsymbol{g}_{\mu\nu} = 8 \pi G \boldsymbol{T}_{\mu\nu} ,
\eeq
assuming that variations of the field are zero at the integration limits (no boundary terms).
For matter fields, consider a combination of a perfect fluid in  equation\,\eqref{eq:Tdiag} and 
an effective minimally coupled scalar field $\psi=\psi(\boldsymbol{x}_{\rm \alpha})$, sometimes called quintessence, with:

\beq
 {\cal L} =  {\cal L}_{\rm m} + {\cal L}_{\rm \psi} \,,
 \eeq
 where
 \beq
 {\cal L}_{\rm \psi} = {1\over{2}} \bar\nabla^2 \psi - V(\psi) \,,
 \label{eq:L_psi}
\eeq
and ${\cal L}_{\rm m}$ is the standard matter\textendash energy content.
We have defined $\bar\nabla^2 \psi \equiv \upartial_\alpha \psi \upartial^\alpha \psi$ and $V(\psi)$ is the potential of the classical scalar field $\psi$. We will next explore the regime where  ${\cal L}$ is dominated by ${\cal L}_{\rm \psi}$. If both ${\cal L}_{\rm m}$ and ${\cal L}_{\rm \psi}$ 
contributions are not coupled, then the general result would correspond to just adding both contributions to $P$ and $\rho$.
We can estimate $\boldsymbol{T}_{\mu\nu}(\psi)$ for $\psi$ from equation\,\eqref{eq:TmunuL}
\beq
\boldsymbol{T}_{\mu\nu}(\psi) =  \upartial_\mu \psi \upartial_\nu \psi - \boldsymbol{g}_{\mu\nu} \left[ {1\over{2}}\bar\nabla^2 \psi - V(\psi) \right] \,.
\label{eq:TmunuPsi}
\eeq
Comparing to a perfect fluid in equation\,\eqref{eq:Tdiag} we can identify (see also equations B66\textendash B68 in  \citealt{Cosmology_book_Weinberg}):
\bea
\rho & =&    {1\over{2}}\bar\nabla^2 \psi  + V(\psi) , \\
P & =&  {1\over{2}}  \bar\nabla^2 \psi- V(\psi) , \\
\boldsymbol{u}_\mu &=&  {\upartial_\mu \psi \over{ \bar\nabla \psi}} ,
\eea
which fulfill $\boldsymbol{u}_\mu \boldsymbol{u}^\mu=1$.
Choosing an observer moving with the fluid corresponds to $\boldsymbol{u}_i=0$ and therefore $\boldsymbol{u}_0 \boldsymbol{u}^0=1$, which is equivalent to:
\bea
\boldsymbol{u}_i=0 ~ \rightarrow ~ \upartial_i \psi &=&  0  , \\
\boldsymbol{u}_0 \boldsymbol{u}^0=1~ \rightarrow ~ \upartial_0 \psi ~\upartial^0 \psi &=& \bar\nabla^2 \psi \,.
\eea
In this frame, we have:
\bea
\rho & =&    {1\over{2}} \boldsymbol{g}^{00} \dot\psi^2 + V(\psi) \,, \label{eq:relat1}\\
P & =&  {1\over{2}} \boldsymbol{g}^{00} \dot\psi^2  - V(\psi) \,, 
\label{eq:relat2}
\eea
where we have defined $\dot\psi \equiv \upartial_0 \psi$.

The ground state of the system corresponds to the configuration in which the energy is minimized. In a relativistic context, this often means that the kinetic contributions (derived from the gradient terms \(\upartial_\mu \psi\)) become insignificant compared to the potential energy \(V(\psi)\):
\beq
V \gg \dot\psi^2\,.
\label{eq:ground-state}
\eeq
This simply means that the energy is dominated by the potential energy (e.g., like it happens to us on the surface of Earth). We expect something similar to happen when we reach the ground state at some supranuclear densities in a cold collapse. The dynamics will be dominated by the potential for interaction of quantum particles, and their kinetic energy will not play a significant role in the evolution at the bounce.
Actually, such a state is also usually assumed to happen during Cosmic Inflation or for DE as given by Quintessence (see, e.g. \citealt{Dodelson}). In this case, we see from equations\,\eqref{eq:relat1} and \eqref{eq:relat2} that $\rho = -P$ and the EoS of the fluid $P=P(\rho)=\omega \rho$ is characterized by $\omega =-1$.

An alternative but more complex approach involves scalar fields with multiple false vacua, where a bubble with a false vacuum interior is created within a region of lower vacuum state (e.g. \citealt{BH_interior_daghigh}). Here, a vacuum refers to a minimum in the potential \(V(\psi)\). Constructing a consistent GR solution for such a configuration requires addressing the issue of discontinuity at the surface that separates the two vacua.

Particularly intriguing are the "Baby Universe" and "Gravastar" solutions, where the interior of a BH is modelled with a de Sitter metric (e.g., \citealt{BH_interior_Gonzalez,1987PhRvD..35.1747B,1989PhLA..138...89G,1989PhLB..216..272F,Aguirre,Farhi,gravastar2015,PBH3}). These solutions are characterized by their inherent discontinuities, necessitating additional matter\textendash energy content at the boundary surface (the bubble), which complicates the situation further. 

It is important to note that these Baby Universes and Gravastar are distinct from the BHU model. Unlike these models, the BHU does not require the introduction of new scalar fields \(\psi\) or potentials \(V(\psi)\), and it avoids the discontinuities or surface terms associated with Baby Universes. The BHU model simply describes an FLRW cloud with a fixed mass \(M_{\rm T}\), which provides a solution to an empty exterior space (\citealt{Gaztanaga:bhu1}). As detailed in Appendix \ref{app:boundary}, while in certain limits, both DE and BHU scenarios can lead to an effective \(\Lambda\) term, the underlying physics behind these terms is quite different. In our paper, we use this ground-state analogy with $\psi(r)$ primarily to understand how the relativistic bouncing solution emerges within the BHU model.

\section{THE GHY BOUNDARY AND THE $\Lambda$ TERM}
\label{app:boundary}

For more than thirty years, cosmologists have accumulated compelling evidence that cosmic expansion is accelerating: \(\ddot{a} > 0\). This acceleration appears to be dominated by a Cosmological Constant term, commonly denoted as \(\Lambda\).

This \(\Lambda\) term can be understood in three distinct ways: (1) as a fundamental (or effective) modification of Einstein's classical GR, denoted here \(\Lambda_{\rm F}\), (2) as an effective source term from the ground state, \(\Lambda_{\rm G}\), which is similar to DE or a Cosmic Inflation component, or (3) as a boundary term, \(\Lambda_{\rm B}\). These three possible origins can be illustrated by writing the Einstein\textendash Hilbert action in equation\,\eqref{eq:actionC} with the corresponding additional terms:
\begin{equation}
    S = \int_{\calMa} \text{d}\calMa \left[ \frac{R}{16\pi G} + {\cal L} - \frac{\Lambda_{\rm F}}{8\pi G} \right] + \frac{1}{8\pi G} \oint_{\upartial \calMa} dV_3 \,K^\prime.
    \label{eq:actionC2}
\end{equation}
The first two terms in the first integral represent the classical GR Lagrangian with matter\textendash energy content \({\cal L}\) as a source term. The third term corresponds to the fundamental cosmological constant \(\Lambda_{\rm F}\). The corresponding equations\,\eqref{eq:H2} and \eqref{eq:ddota} have to include new terms containing \(\Lambda_{\rm F}\):
\bea
\left[\frac{\dot{a}}{a}\right]^2 + \frac{k}{a^2} &=& \frac{8\pi G}{3} \rho + \frac{\Lambda_{\rm F}}{3}\,, \label{eq:dotaLambda} \\
\frac{\ddot{a}}{a} &=& - \frac{4\pi G}{3} (\rho + 3P) + \frac{\Lambda_{\rm F}}{3}\,,
\label{eq:ddotaLambda} 
\eea
where \(P\) and \(\rho\) can have contributions from different source components. These equations are the same for the finite FLRW cloud. We can express \({\cal L}\) as given by equation\,\eqref{eq:L_psi} where \({\cal L}_{\rm \psi} = \frac{1}{2} \bar{\nabla}^2 \psi - V(\psi)\). As shown in the previous section, when the sources are dominated by a constant ground state (G), $\psi_{\rm G}$, this term reduces to ${\cal L}_{\rm \psi} \simeq {\cal L}_{\rm G} = P_{\rm G} = -\Lambda_{\rm G} = -V(\psi_{\rm G})$ being constant, see equations \eqref{eq:relat1} and \eqref{eq:relat2}, and we can define:
\beq
\Lambda_{\rm G} \equiv 8\pi G \rho_{\rm G} = 8\pi G V(\psi_{\rm G}).
\eeq
In this situation, we have \({\cal L} \simeq \frac{\Lambda_{\rm G}}{8\pi G}\) in equation\,\eqref{eq:actionC2}, so that \(\Lambda_{\rm F}\) and \(\Lambda_{\rm G}\) provide degenerate interpretations for \(\Lambda\) in the context of cosmology. In equations\,\eqref{eq:ddotaLambda} and \eqref{eq:dotaLambda}, we can see both terms separated ($\Lambda_{\rm G}$ appears as a contribution $\rho \simeq -P \simeq \frac{\Lambda_{\rm G}}{8\pi G}$), so they represent different possible changes to the action, which could emulate the effect of the other. In the limit where \(\Lambda_{\rm G}\) is constant, we cannot tell if \(\Lambda\) is DE, a ground state (i.e., vacuum energy), or a fundamental modification to GR. However, \(\Lambda_{\rm G}\) can be dynamical and could appear or become negligible compared to the dynamic components (as we see in our bouncing scenario, in Cosmic Inflation and possibly for DE), while \(\Lambda_{\rm F}\) is usually assumed to be a constant (but can also be dynamical in more elaborate modified gravity models).

The last integral in equation\,\eqref{eq:actionC2} represents the Gibbons\textendash Hawking\textendash York (GHY) boundary term \citep{York, Gibbons, Hawking1996}, where \( K^\prime \) is the trace of the extrinsic curvature at the boundary \( \upartial \calMa = V_3 \). As shown in \cite{gaztanaga_mou}, for the FLRW cloud with total mass\textendash energy \( M_{\rm T} \), this term results in:
\bea
    S_{\rm GHY} &=& \frac{1}{8\pi G} \oint_{\upartial \calMa} dV_3 \,K^\prime = \int_\calMa \text{d}\calMa \left[ \frac{-2\Lambda_{\rm B}}{16\pi G} \right]\,, \\
    \Lambda_{\rm B} &\equiv&  \frac{3}{r_{\rm s}^2} \quad \text{where} \quad r_{\rm s}= 2GM_{\rm T} \, .
\eea
Equations\,\eqref{eq:ddotaLambda} remains the same with $\Lambda_{\rm F}$ replaced by $\Lambda_{\rm B}$:
\bea
\left[\frac{\dot{a}}{a}\right]^2 + \frac{k}{a^2} &=& \frac{8\pi G}{3} \rho + \frac{1}{r_{\rm s}^2}, \nonumber \\
\frac{\ddot{a}}{a} &=& - \frac{4\pi G}{3} (\rho + 3P) + \frac{1}{r_{\rm s}^2},
\label{eq:ddotaLambda2}
\eea
This provides a fundamentally different origin for \(\Lambda\) compared to \(\Lambda_{\rm F}\) or \(\Lambda_{\rm G}\): it corresponds to a finite-mass \( M_{\rm T} \) FLRW cloud trapped within its own gravitational radius \( r_{\rm s} = 2GM_{\rm T} \). This boundary interpretation for the observed \(\Lambda\) term corresponds to the BHU model.

In the bouncing scenario presented in the main text of this paper, we start with a large (\( R > r_{\rm s} \)) and low-density cloud with \(\Lambda_{\rm F} = \Lambda_{\rm G} = \Lambda_{\rm B} = 0\), with \(\rho \neq 0\) and \( P \simeq 0 \). As shown in equation\,\eqref{eq:E=0}, this can be modelled with a Newtonian equivalent problem, which also does not have any \(\Lambda\) term. In the Newtonian analogue, the collapse is halted by the polytropic EoS, which (as we showed here in Fig.\,\ref{fig:bounce9}) acts similarly to the GR ground state term, with negative pressure \( P_{\rm G} = -\rho_{\rm G} \) and positive curvature \( k \). In the above equations, the following bounce would act in a similar way as the \(\Lambda_{\rm G}\) term, which produces the exponential expansion (the analogue of a Supernova explosion). But once the bounce occurs, the density and pressure reduce again (because the system is no longer in a ground state), and \(\Lambda_{\rm G}\) disappears. The curvature \( k \) is also diluted away by the initial exponential expansion. When the bounce occurs inside the BH, we need to add the GHY boundary term, \(\Lambda_{\rm B} = \frac{3}{r_{\rm s}^2}\), and this prevents the expansion from escaping the gravitational radius \( r_{\rm s} \). We explain how this term prevents the White Hole solution in Section \ref{sec:WH}. Thus, \(\Lambda_{\rm B}\) and therefore \( M_{\rm T} \), prevent matter from escaping outside \( r_{\rm s} \).

\end{document}